\documentclass[preprint,proceedings,useAMS,usenatbib]{rmaa}
\usepackage{epsfig,rotating,natbib,psfrag}
\def\msun{${\rm M_{\odot}} \;$}
\def\Msun{${\rm M_{\odot}}$}
\newcommand\Mesz{M\'esz\'aros~}
\newcommand\Pacz{Paczy\'nski~}

\newcommand\be{\begin{equation}}
\newcommand\ee{\end{equation}}
\newcommand\bi{\begin{itemize}}
\newcommand\ii{\item}
\newcommand\ei{\end{itemize}}
\newcommand\bea{\begin{eqnarray}}
\newcommand\eea{\end{eqnarray}}
\newcommand\gcc{gcm$^{-3}$}

\SetYear{2005}
\SetConfTitle{Cozumel}

\title{Last moments in the life of a compact binary system: \\
gravitational waves, gamma-ray bursts and magnetar formation}

\author{S. Rosswog\altaffilmark{1}}

\altaffiltext{1}{School of Engineering and Science, International University
  Bremen, Germany; Jacobs University Bremen as of spring 2007.}

\shortauthor{Rosswog}%
\shorttitle{Last moments}

\fulladdresses{
\item Stephan Rosswog: School of Engineering and Science,
  International University Bremen, Campus Ring 1, D-28759 Bremen, Germany
  (\email{s.rosswog@iu-bremen.de}).   
}

\listofauthors{S. Rosswog}
\indexauthor{Rosswog, S.}

\abstract{The first detections of afterglows from {\em short} gamma-ray
  bursts (GRBs) have confirmed the previous suspicion that they are 
  triggered by a different central engine
  than long bursts. In particular, the recent detections of short GRBs in 
  galaxies without star formation lends support to the idea that an old
  stellar population is involved. Most prominent are 
  mergers of either double neutron stars or of a neutron star
  with a stellar-mass black hole companion. Since the final identification of
  the central engine will only come from an integral view of several
  properties, we review the observable signatures that can be expected from
  both double neutron stars and neutron star black hole systems. We discuss
  the gravitational wave emission, the structure of the neutrino-cooled
  accretion disks, the resulting neutrino signal and possible mechanisms to 
  launch a GRB. In addition, we address the speculative idea that in
  some cases a magnetar-like object may be the final outcome of a double
  neutron star merger. We also discuss possibilities to explain the 
  late-time X-ray activity that has been observed in several bursts.}

\addkeyword{gravitational waves}
\addkeyword{jets and outflows}
\addkeyword{neutrino-cooled disks}
\addkeyword{gamma-ray bursts}
\addkeyword{magnetars}

\begin{document}
% Typeset article header
\maketitle
\section{Introduction}
Gamma-ray bursts (GRBs) are subdivided on grounds of
both their durations and spectra into short-hard and long-soft bursts
(Kouveliotou et al. 1993).  While a lot has been learned from the 
detection of long GRB afterglows since 1997 (Costa et al. 1997, van Paradijs
et al. 1997, Frail et al. 1997, Bloom et al. 1999, Stanek et al. 2003, Hjorth
et al. 2003), for example that long GRBs are related to death of massive stars
in star-forming regions, the afterglows of short GRBs remained elusive 
until the early summer of 2005. The recent detections of afterglows from short
GRBs point to a central engine that is related to an old stellar
population. In particular, it has been argued (Bloom et al. 2006, Berger et
al. 2005, Barthelmy et al. 2005, Villasenor et al. 2005) that the observations
are consistent with being the result of compact binary mergers: they occur
systematically at lower redshifts\footnote{It has recently been suggested 
(Berger et al. 2006b) that
  at least 1/4 of the short bursts could lie at redshifts $z>0.7$ and would
  therefore imply substantially larger isotropised energies than was inferred
  for the first set of detected short GRBs with afterglows.} than their 
long-duration cousins (e.g. Fox
et al. 2005), both in galaxies with (Hjorth et al. 2005) and
without star formation (Berger et al. 2005, Barthelmy et al. 2005) and they
are not accompanied by a detectable supernova explosion (Bloom et al. 2006, 
Fox et al. 2005, Hjorth et al. 2005).\\ 
\begin{figure}[!t]
%low-res versions
 \includegraphics[angle=-90,width=0.75\columnwidth]{dens_cut_fd36.ps}
  \includegraphics[angle=-90,width=0.75\columnwidth]{dens_cut_fd45.ps}
  \includegraphics[angle=-90,width=0.75\columnwidth]{dens_cut_fd120.ps}
 \includegraphics[angle=-90,width=0.77\columnwidth]{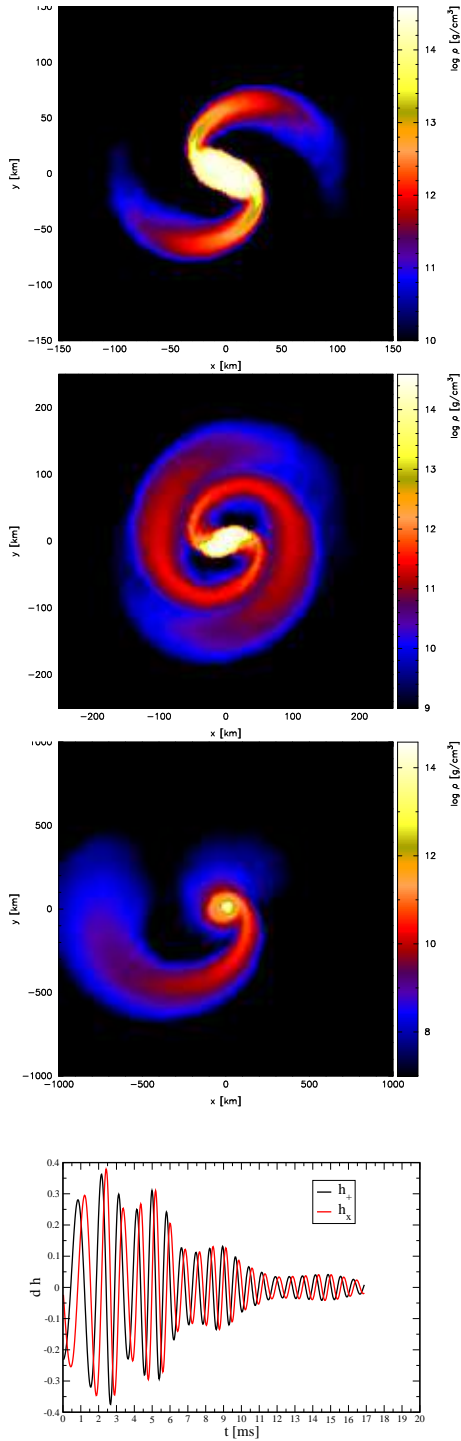}
  \caption{Merger of a double neutron star system with 1.4 \msun
  each (snapshots at 4.41, 5.54 and 15.0 ms). 
%Color-coded is density in the orbital plane at t= 4.41, 5.54 and 15.0 ms. 
The lowest panel shows the gravitational wave amplitudes times the distance
to the source, $d$ (in units of 1.5 km).}
  \label{fig:DNS_1.4}
\end{figure}
There are, however, puzzling observations of late-time flaring activity in
several short bursts Berger et al. 2006a). For example, GRB050724 showed long-lasting ($\sim 100
$s) X-ray flaring activity after a delay of $\sim 30$ s. These long-lasting
flares motivated MacFadyen et al. (2006) to suggest an accretion-induced
collapse of a neutron star in a low-mass X-ray binary as an alternative 
central engine. This central engine would produce the flaring by interaction 
of the GRB outflow with the non-compact companion star.\\ 
As the final identification of the central engine will probably only come from
an integral view of various features from a large sample of events, we will
give an overview over the properties and signatures that can be expected from
compact binary mergers. They will include gravitational waves, the structure of
the forming accretion disk, the neutrino emission and the burst
itself. Moreover, we will discuss possibilities to produce the observed
late-time flaring activity.\\ 
The discussion is mostly based on own simulations, some of which have been
reported previously  (Rosswog and Davies 2002, Rosswog and Ramirez-Ruiz 2002
and 2003, Rosswog and Liebend\"orfer 2003, Rosswog et al. 2003, Rosswog et
al. 2004, Rosswog 2005) and several new simulations that are introduced here, 
see Table \ref{runs}. The physics included in our models and the numerical 
techniques that we use have been described in a series of papers (Rosswog 
and Davies 2002, Rosswog and Liebend\"orfer 2003, Rosswog et al. 2003, 
Rosswog et al. 2004).
\begin{figure}
\includegraphics[angle=-90,width=0.75\columnwidth]{dens_cut_ns11_ns16_fd13.ps}
\includegraphics[angle=-90,width=0.75\columnwidth]{dens_cut_ns11_ns16_fd50.ps}
\includegraphics[angle=-90,width=0.75\columnwidth]{dens_cut_ns11_ns16_fd120.ps}
\includegraphics[angle=-90,width=0.77\columnwidth]{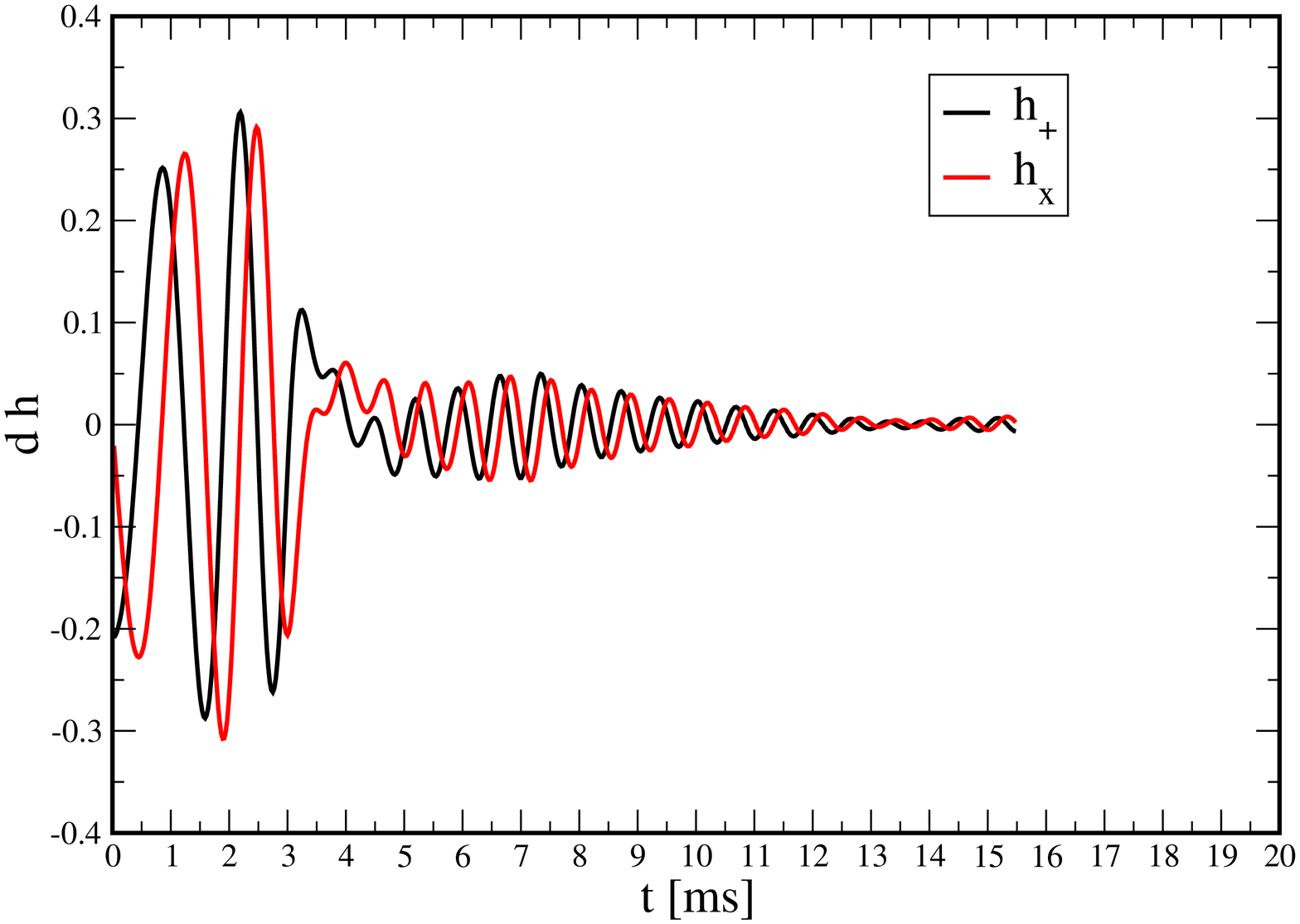}
  \caption{DNS merger with 1.1 and 1.6 \msun (at 1.51, 6.17 and 15.0 ms).
 The lowest panel shows the gravitational wave amplitudes times the distance
to the source, $d$ (in units of 1.5 km), as a function of time.} 
  \label{fig:DNS_1.1_1.6}
\end{figure}
\begin{figure}
\includegraphics[angle=-90,width=0.99\columnwidth]{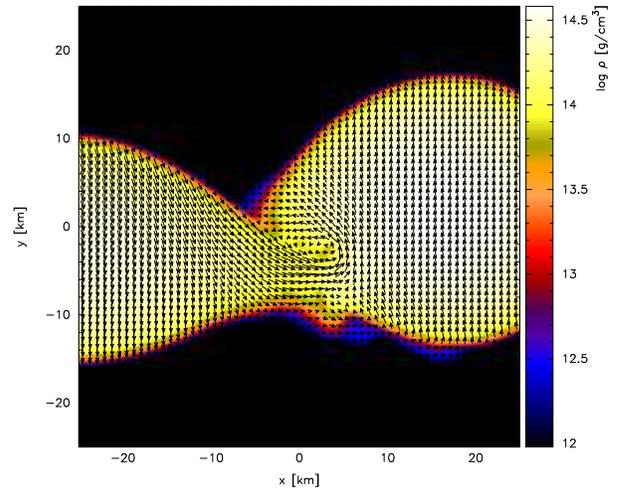}

  \caption{Merger of 1.1 and 1.6 \msun system (t=1.63 ms): Impact of accretion
    stream onto 
    surface of accreting neutron star. Color-coded is density in the orbital
  plane, the arrows indicate velocities.} 
  \label{fig:algol_impact}
\end{figure}

The new neutron star merger calculations have been produced with a new Smoothed
Particle Hydrodynamics (SPH) code that benefits from a slew of numerical
improvements. It incorporates 
\bi
\ii an artificial viscosity oriented at Riemann solvers, 
    see Chow and Monaghan (1997),
\ii a consistent accounting of the effects of the so-called ``grad-h''-terms, 
    see Springel and Hernquist (2002), Monaghan (2002) and Price (2004),
\ii a consistent implementation of adaptive gravitational softening 
    lengths, see Price and Monaghan (2006),
\ii the option to  evolve magnetic fields together with the fluid, 
    either via so-called Euler potentials (Stern 1970) or via a SPH 
    discretisation of the MHD-equations (Price and Monaghan 2005).
\ei
First results obtained with this code have been published (Price and Rosswog
2006), a detailed code documentation can be found in Rosswog and Price (2006).
 
%\section{Which physics is important?}

\section{Gravitational waves as probes of the coalescence dynamics}
\label{sec:dynamics} 

\begin{table*}
\begin{flushleft}
\caption{New runs: DNS stands for double neutron star systems, NSBH for
  neutron star black hole systems. M$_1$ and M$_2$ are the component masses, 
  $q$ is the mass ratio,
  a$_0$ is the initial separation, N stands for Newtonian, PW for the
  \Pacz-Wiita pseudo-potential, $R_{\rm abs}$ is the radius of the absorbing 
  boundary, in units of $G M_{\rm bh}/c^2$, the next column gives the SPH particle number.} \label{runs}
\begin{tabular}{rccccccrccccccccc} 
run & type & M$_1$, M$_2$, $q$ & a$_0$ [km] & grav. & R$_{\rm abs}$ & SPH  part. &\\ \hline \\
DA  & DNS  & 1.4,  1.4 ,  1      & 48    & N  & - &$2\cdot 10^6$ &\\
DB  & DNS  & 1.1,  1.6 ,  0.6875 & 48    & N  & - &$5\cdot 10^5$ &\\
NA  & NSBH & 1.4,  3.0 ,  0.4667 & 60    & N  & 6 &$6\cdot 10^5$ &\\
NB  & NSBH & 1.4,  4.0 ,  0.35   & 72    & N  & 6 &$2\cdot 10^5$ &\\
NC  & NSBH & 1.4,  7.0 ,  0.2    & 90    & N  & 6 &$2\cdot 10^5$ &\\
ND  & NSBH & 1.4, 10.0 ,  0.14   & 112.5 & PW & 3 &$2\cdot 10^5$ &\\
\end{tabular}
\end{flushleft}
\end{table*} 

%GWs as probes
Gravitational waves can serve as a direct probe of the merger dynamics and, if
detected coincident with a short GRB, they would provide the ultimate proof of
the compact binary nature of the central engine. The dynamics of the
merger process is sensitive to the mass ratio of 
the involved components, therefore, the signal from the merger of a double 
neutron star system (DNS) with a mass ratio close to unity can be very 
different from a neutron star black hole (NSBH), where the black hole can, 
in principle, be much more massive than the neutron star. \\
%GW detection
In the last minutes before the coalescence the gravitational wave signal
will slowly sweep through the frequency range that is accessible to
ground-based gravitational wave-detectors 
such as GEO600 (Grote et al. 2005), LIGO (Abramovici et al. 1992), TAMA (Ando
et al. 2004) or VIRGO (Spallici et al. 2005, Freise 2005). 
The detection of a gravitational wave signal coincident with a short GRB would
allow to determine the distance to the source, the luminosity and the beaming
angle (Kobayashi and \Mesz 2003).\\
%The expectations are about one
%double neutron star detection per about 30 years for LIGO I and about one per
%5 years for advanced LIGO (Kalogera et al. 2004).\\
The coalescence of a double neutron star system
can be divided into three phases: the secular inspiral due to
gravitational wave emission, the actual merger and, finally, the ``ring-down''
phase, in which the death-struggle of the freshly formed super-massive neutron
star will take place. After this phase the remnant will have settled into its
final state, probably a black hole (an alternative is discussed in
Sec.~\ref{sec:magnetar}).\\  
During the inspiral, the gravitational wave forms can be accurately described 
via post-Newtonian expansions for point masses. To date, post-Newtonian
formalisms exist up to 3.5PN order, see Blanchet (2006) for a recent
review, and lowest order spin-spin and spin-orbit couplings can be accounted
for (e.g. Will 2005).  In the
inspiral phase, both the frequency and the amplitude of the waves will 
increase, the system is said to ``chirp''. The orbit decays secularly until 
the last stable orbit is reached, where the binary enters a ``plunging
phase''. In this phase the stars fall nearly radially towards each other 
and merge within about one orbital period. This dynamical instability 
(Chandrasekhar 1975, Tassoul 1975) is the result a steepening of the 
gravitational potential due to both purely Newtonian tidal (Lai et al. 1993) 
and general relativistic effects, for a recent review on relativistic binaries
see Baumgarte and Shapiro (2003).  For the last two phases, the merger and 
ringdown, three-dimensional hydrodynamic simulations are required to 
predict the gravitational wave signal.\\ 
%
%DNS
%
%work in this field
Double neutron star merger wave forms has been predicted by several groups. 
%'Newtonian'
The calculations started with Newtonian calculations
(e.g. Oohara and Nakamura 1989, Rasio and Shapiro 1994, Ruffert et al. 1997,
2001).
%Post-Newtonian
Post-Newtonian approaches (Ayal et al. 2001, Faber and Rasio 2000) 
turned out not to be particularly successful due to the importance of 
higher order PN-corrections.
%Conformal flatness
In the conformal flatness approach (Isenberg 1978, Wilson et al. 1996) it
is assumed that the dynamical degrees of freedom of the gravitational
fields, i.e. gravitational waves, can be neglected and that the spatial 
part of the metric is (up to a conformal factor) flat and remains so during the
further evolution. This approximation has been used in several compact binary
simulations (e.g. Wilson et al. 1996, Oechslin et al. 2002, 2004, 
Faber et al. 2004, Oechslin 2006) and it could be shown (Cook et al. 1996)
to be provide reasonable accuracy in several cases.
%Full GR
Finally, fully relativistic simulations have been performed by Shibata and
collaborators (Shibata 1999, Shibata and Uryu 2002, Shibata 
et al. 2005). For a recent review on numerical relativity
and compact binaries we refer to Baumgarte and Shapiro (2003).\\
%
%NSBH
%
The simulation of neutron star black hole binaries has been somewhat lagging
behind the DNS status. Again, first simulations were  Newtonian
(e.g. Lee 1999a,b, Janka 1999, Lee 2000, 2001, Rosswog et al. 2004), slightly
later followed simulations that used a \Pacz-Wiita (1980) pseudo-potential
(Lee and Kluzniak 1999a, Rosswog 2005, Setiavan et al. 2004, 2006).
Very recently, progress has been made with relativistic approaches. 
For example, Taniguchi et al. (2005) were able to construct general 
relativistic quasi-equilibrium sequences for black holes that are much 
more massive than the neutron star. Faber et al. (2006) used an exact metric
together with a conformal flatness approach in which the black hole 
position was artifically kept fixed in space.
L\"offler et al. (2006) were able to treat binary components of comparable
mass, but their approach was restricted to a head-on collision. Recently, 
Shibata and Uryu (2006) reported on first results of neutron star black hole 
binaries in full general relativity.\\
Here, we will report mainly on our own, either Newtonian or pseudo-Newtonian
results. These calculations focused on the microphysics rather than the
strong-field gravity aspect, therefore, the results concerning gravitational
waves should be taken with a grain of salt. Details about the calculation of
the wave forms can be found in Rosswog et al. (2004), we will mention in the
appropriate places in which direction fully relativistic calculations are
expected to change the results. 

\subsection{Double neutron stars}
For many years, neutron star masses, at least those in double neutron star
systems, were thought to be tightly clustered around a mass of  m=
1.35 \msun (Thorsett and Chakrabarty 1999). Recent data, see e.g. Stairs
(2004), show considerable deviations from this value for several of the known 
binary systems. The recently discovered double neutron star system 
PSR J1756-2251 (Faulkner et al. 2005), for example, has a mass ratio of 
q= 0.84 ($m_1= 1.40$ and $m_2= 1.18$ \Msun) and the double neutron star 
J1518+4904 even has $q= 0.67$ ($m_1= 1.56$ and $m_2= 1.05$ \Msun), but 
still relatively large errors (see Stairs 2004).\\ 
For our new DNS simulations, we take masses of twice 1.4 \msun as the 
typical case, but we also consider the more extreme case with a 1.1 and a 1.6
\msun neutron star. The dynamical evolution during the last moments of 
a binary system with twice 1.4 \msun is shown in Fig.~\ref{fig:DNS_1.4}.
The neutron stars have no initial spin, their initial separation is just
outside the plunging regime for our equation of state at 48 km and each star
is modelled with slightly more than a million SPH particles (run DA in 
Tab.~\ref{runs}). Once the plunging sets in, the stars merge within about
one orbital revolution. Excess angular momentum is shed into symmetric spiral
arms that subsequently spread into an extended accretion disk, see 
Sec.~\ref{sec:accretion}, around a super-massive, neutron star-like object.
The evolution of the 1.1 and 1.6 \msun system (run DB in Tab.~\ref{runs}), 
similar to J1518+4904, is shown in Fig.~\ref{fig:DNS_1.1_1.6}. In this case
the lighter star starts to transfer mass towards and then completely
engulfs the more massive component. The mass transfer leads to a complete 
tidal disruption of the 1.1 \msun star, parts of which form a tidal tail, 
see panel three of Fig.~\ref{fig:DNS_1.1_1.6}. For the chosen system 
parameters, the circularization radius of the accreted material is smaller 
than the accretor radius, therefore, like in an Algol system, the accretion 
stream directly impacts on the accretor crust, see panel one in 
Fig.~\ref{fig:DNS_1.1_1.6} and Fig.~\ref{fig:algol_impact}. For less 
extreme mass ratios the accreted material slides more smoothly around the 
surface of the heavier companion, otherwise the evolution is similar.\\
In the last panels of Figs.~\ref{fig:DNS_1.4} and \ref{fig:DNS_1.1_1.6} 
we show the retarded gravitational wave amplitudes, $h_+$ and $h_\times$, 
as seen along the binary axis by a distant observer at distance $d$.
In both cases one sees the last stages of the ``chirp''-signal up to 
about 2.5 ms, when the stars come into contact.
For the chosen initial separation of the 2 x 1.4 \msun case the peak 
frequency of $\approx$ 1 KHz is reached 2.5 ms after the simulation start. 
If $h_{\rm min}$ labels the minimum detectable amplitude, such a system
would be visible out to a distance $d= 18 \; {\rm Mpc} \; 
(10^{-21}/h_{\rm min})$. During the chirp phase up to about
2.5 ms the signal is the result of the binary orbital motion, after that, up 
to about 6 ms it is determined by the elongated central object and the spiral
arms. The central object does not become rotationally symmetric by the end 
of the simulation and, therefore, keeps emitting quasi-periodic 
gravitational waves at about a tenth of the amplitude. This is behaviour
is expected for a stiff equation of state (EOS) such as the 
Shen et al. (1998) EOS that we use. It had been noted very early on (Rasio 
and Shapiro 1994) that this post-merger gravitational wave signal is a way 
to constrain the equation of state.\\
For the 1.1-1.6 \msun system the peak frequency is slightly lower, 
$\nu_{\rm peak} \approx 780$ Hz, the same statement is true for 
the peak amplitude.\\
\begin{figure}
\includegraphics[angle=-90,width=1.1\columnwidth]{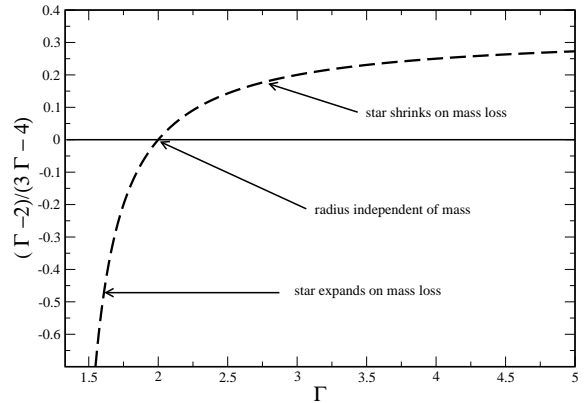}
  \caption{Exponent of the mass-radius relationship for a polytropic star.}  
  \label{fig:m-r-expon}
\end{figure}

\begin{figure*}
\centerline{\includegraphics[angle=-90,width=1.15\columnwidth]{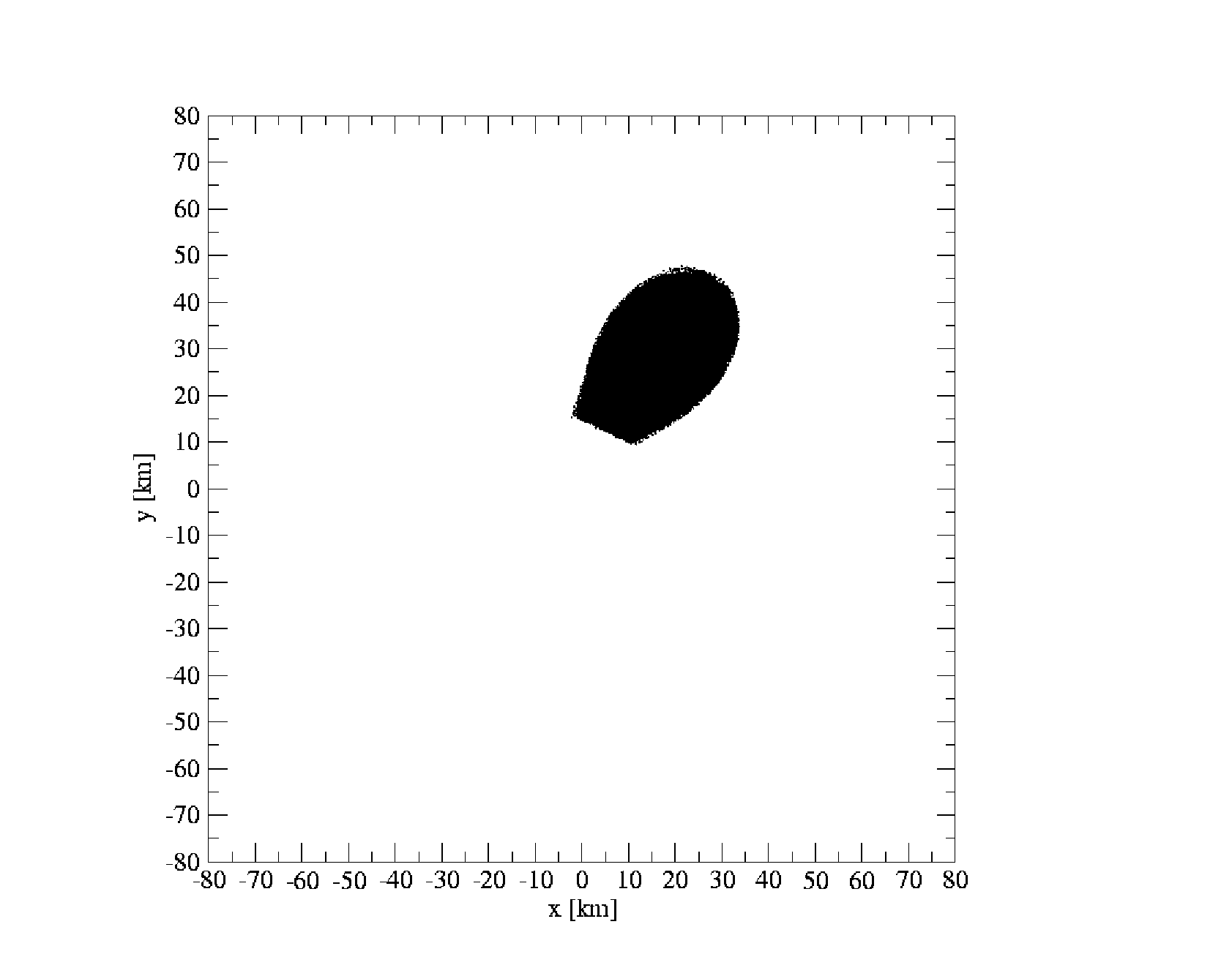}
            \includegraphics[angle=-90,width=1.15\columnwidth]{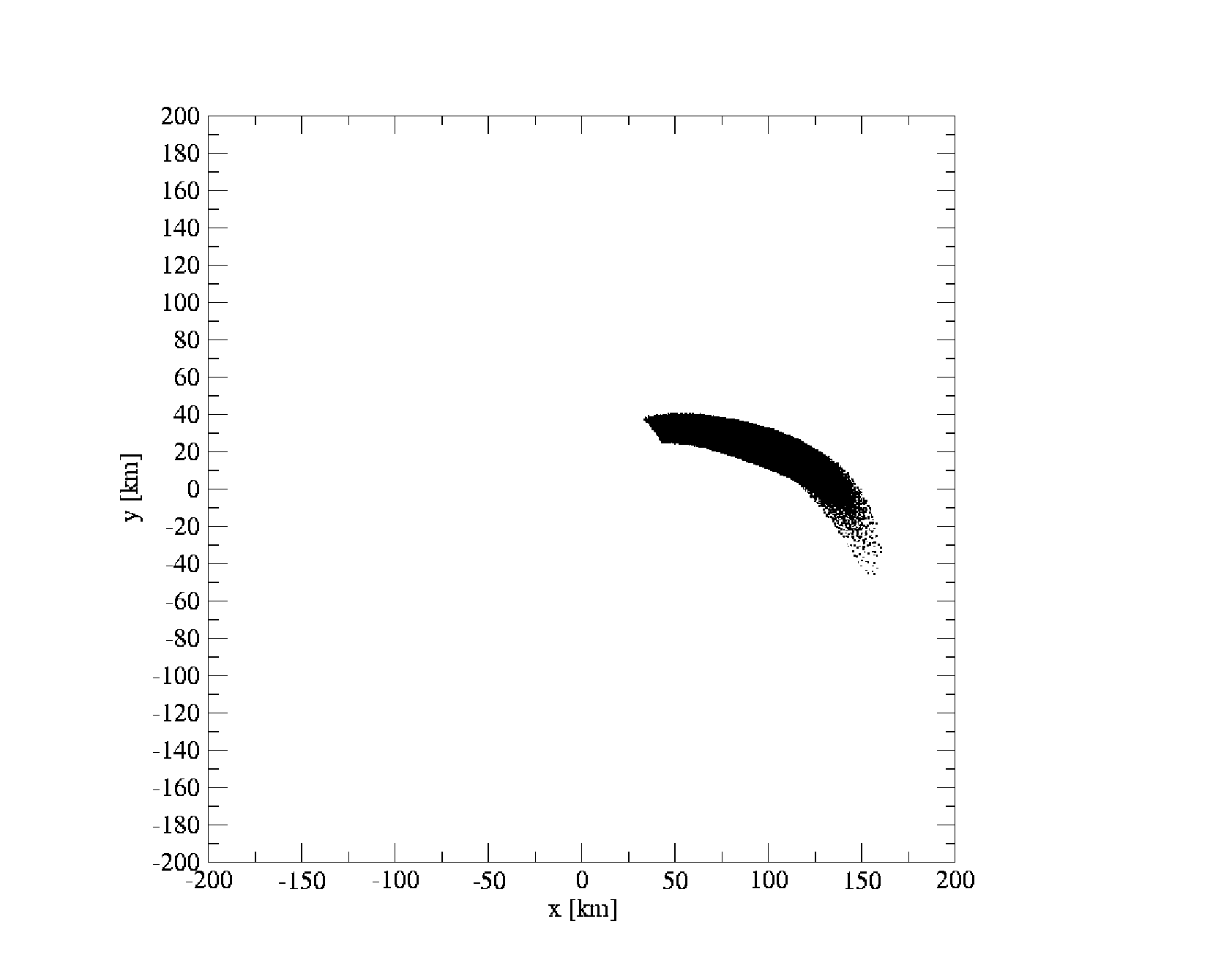}}
\centerline{\includegraphics[angle=-90,width=1.15\columnwidth]{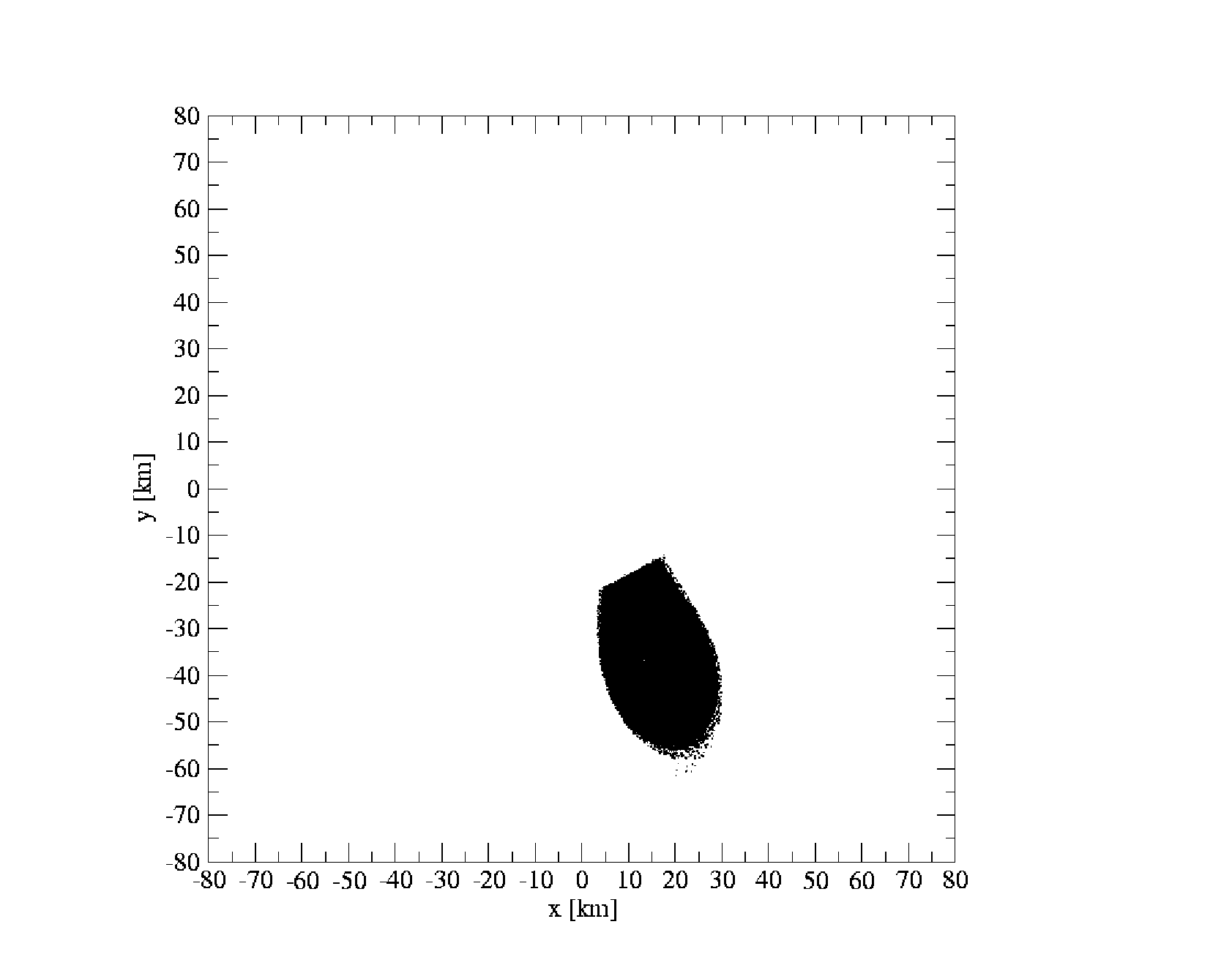}
            \includegraphics[angle=-90,width=1.15\columnwidth]{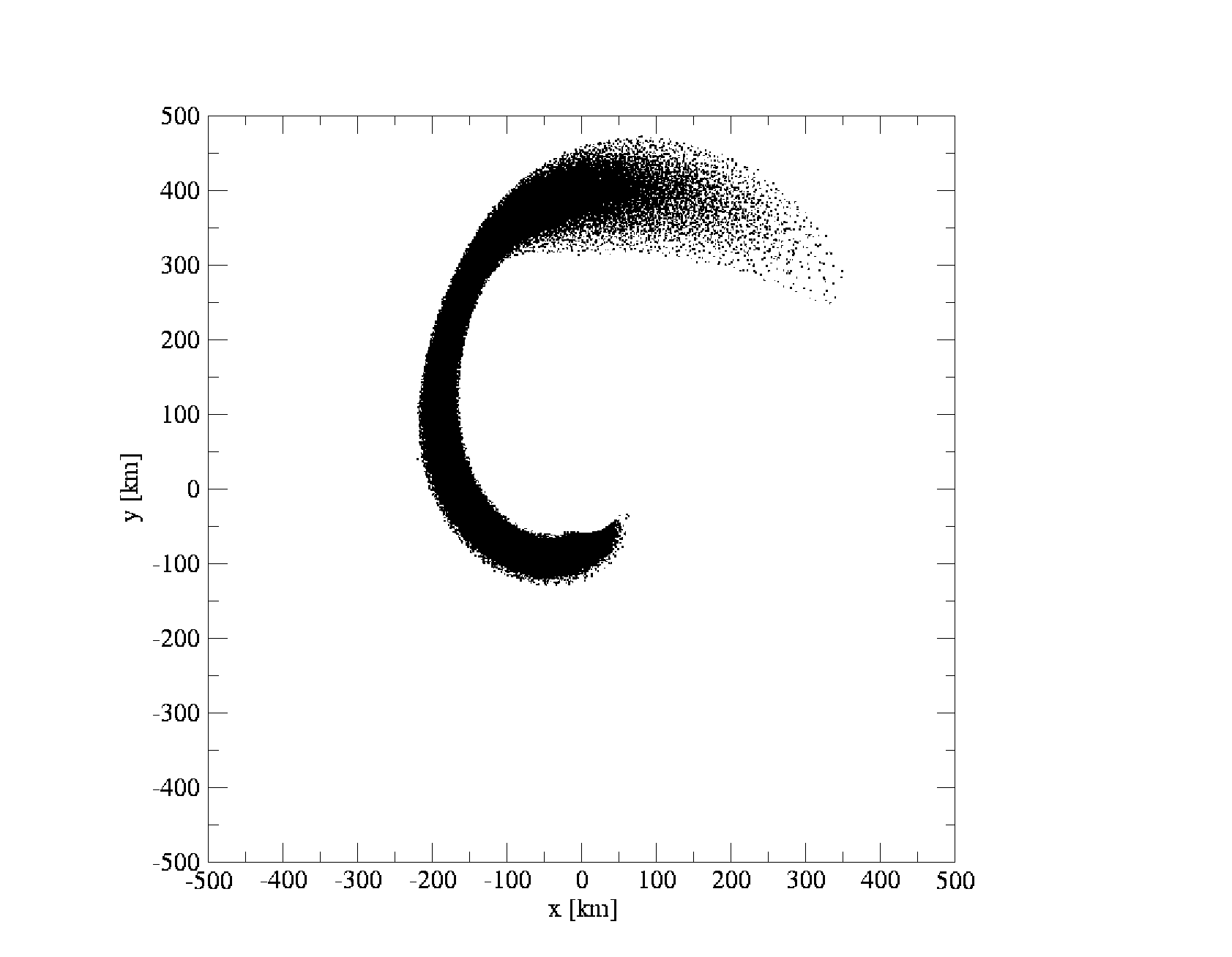}}
\centerline{\includegraphics[angle=-90,width=1.15\columnwidth]{position_fd162_nsbh_q046.ps}
            \includegraphics[angle=-90,width=1.2\columnwidth]{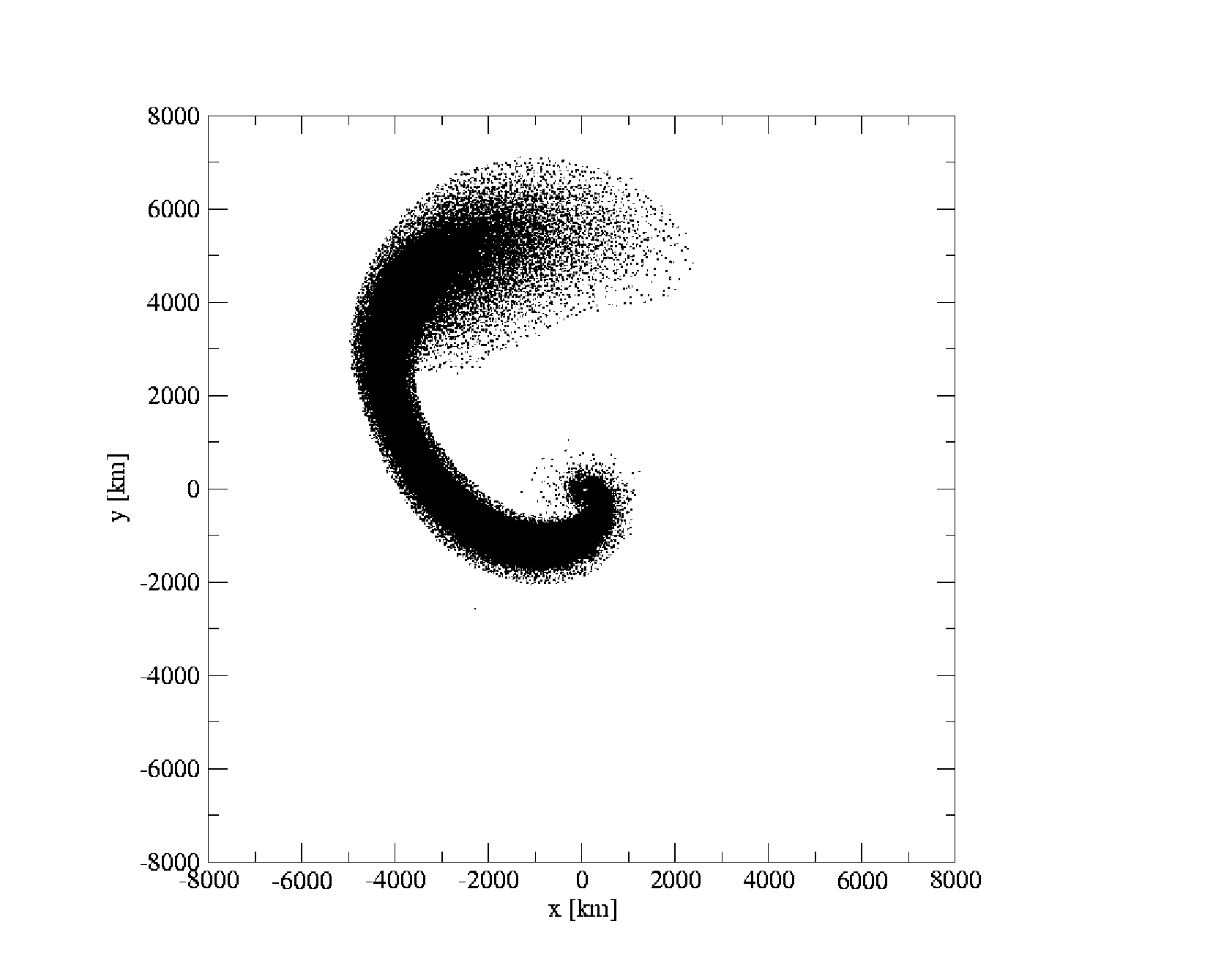}}
  \caption{Dynamical evolution of a neutron star black hole system.
           Left column: example of a system that undergoes episodic mass transfer
          (M$_{\rm ns}$= 1.4 \Msun, M$_{\rm bh}$= 3 \Msun). A mini-neutron star
          keeps orbiting the hole, thereby transfers mass directly into
          the hole and sheds mass which subsequently spreads into a dilute disk
          surrounding the orbiting binary system. It is only after 47 orbital
          revolutions ($t\approx 220$ ms) that the mini-neutron star is finally 
          disrupted, see panel 4, and its remains form an accretion
          disk of $\approx 0.05$ \Msun, see Fig.~\ref{fig:bound_debris}. This
          episodic mass transfer is imprinted on the gravitational wave
          signal, see Fig.~\ref{fig:GW_comparison}.
          Right column: binary in which the neutron star is directly disrupted
          (M$_{\rm ns}$= 1.4 \Msun, M$_{\rm bh}$= 14 \Msun). The figures in
          each column refer to 3, 5 and 37 ms, respectively, after the onset
          of mass transfer.}   
  \label{fig:nsbh_dynamics}
\end{figure*}

\begin{figure}
\centerline{\includegraphics[angle=-90,width=1.15\columnwidth]{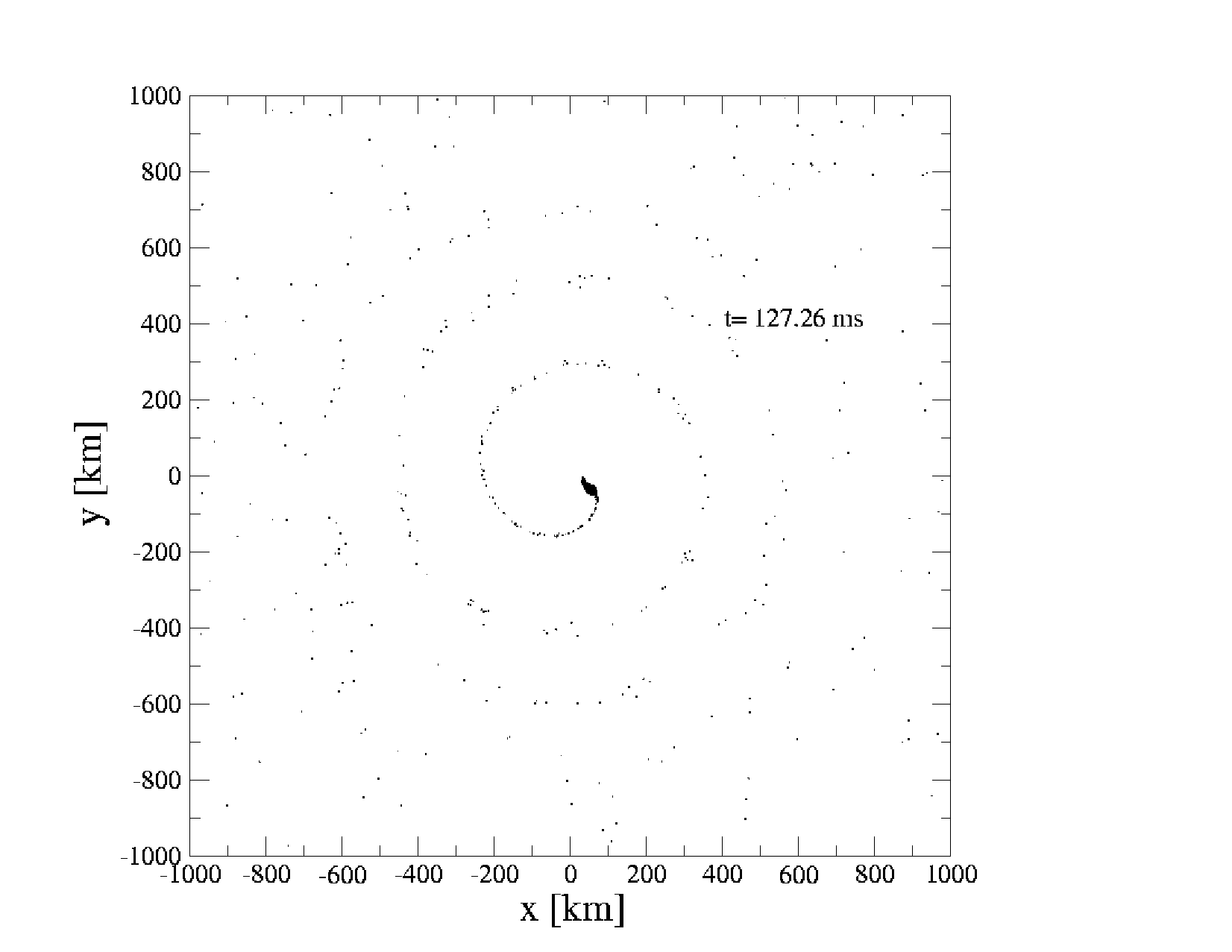}}
\centerline{\includegraphics[angle=-90,width=1.15\columnwidth]{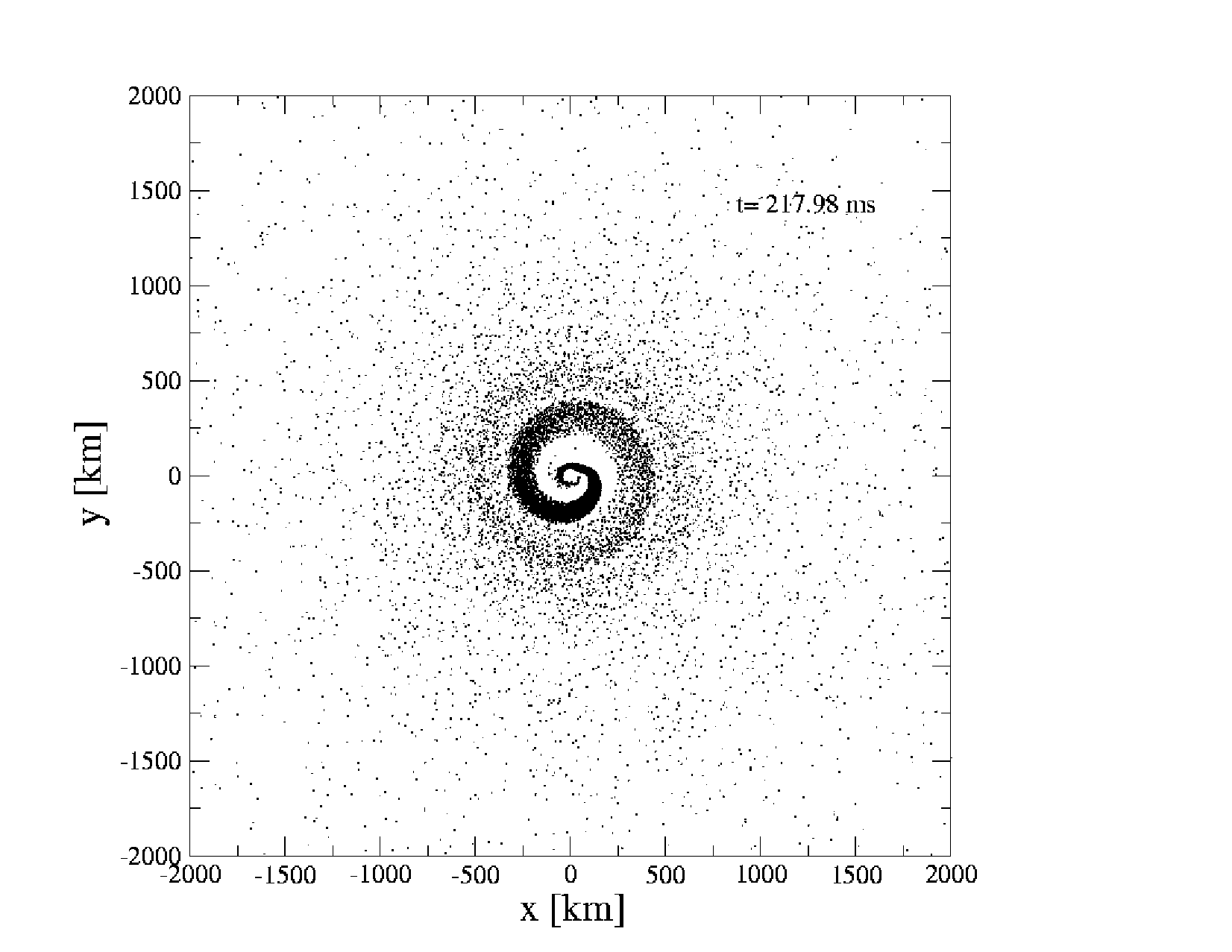}}
\centerline{\includegraphics[angle=-90,width=1.2\columnwidth]{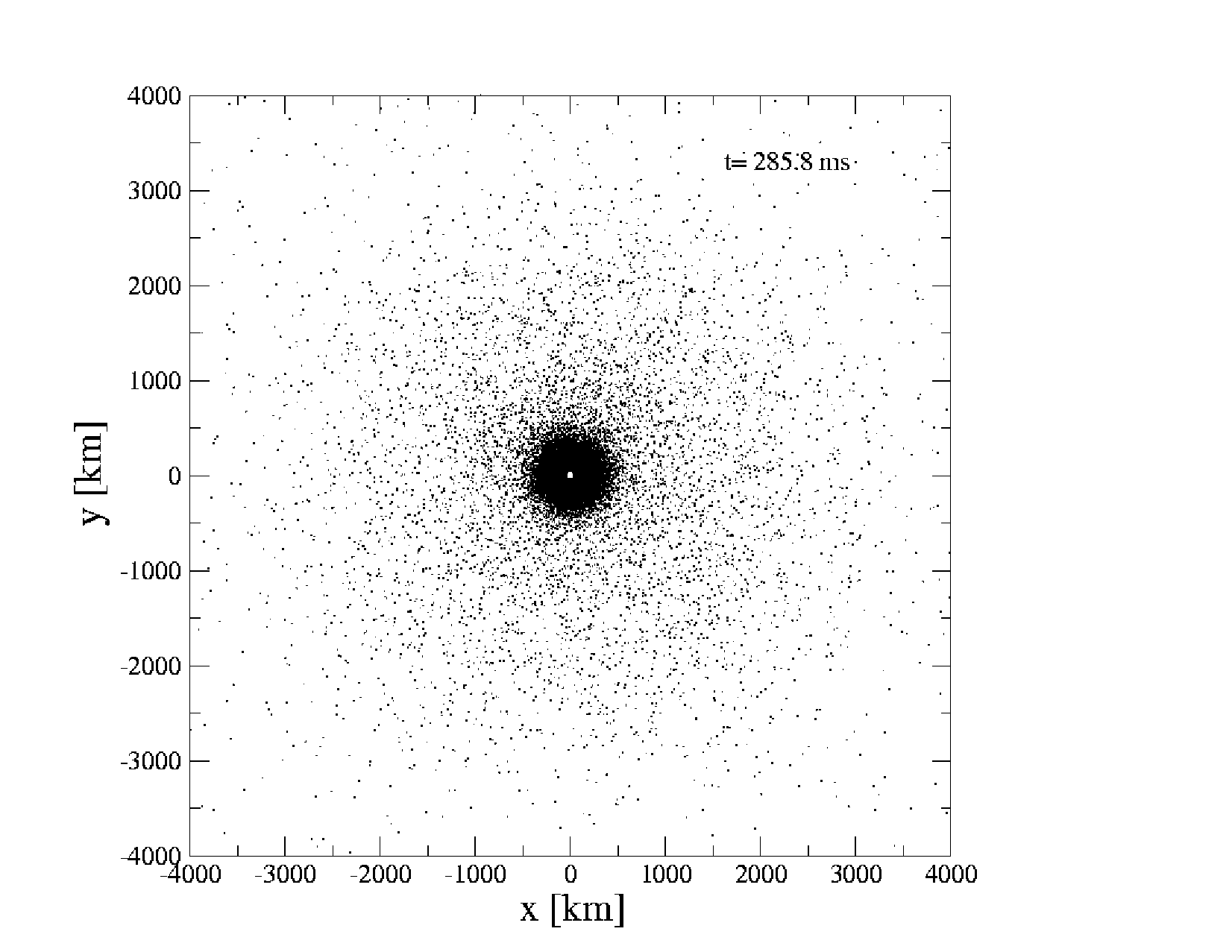}}
  \caption{Dynamical evolution of a neutron star black hole system that
          undergoes episodic mass transfer; continuation of
          Fig.~\ref{fig:nsbh_dynamics}, left column.}   
  \label{fig:nsbh_dynamics_cont}
\end{figure}

This basic picture of the dynamical evolution carries over to general
relativistic calculations. The stronger gravity, however, produces
neutron stars of higher compactness and a more compact remnant. Apart from the
most obvious consequence, the possible gravitational collapse to a black hole,
a general relativistic merger yields higher peak frequencies and gravitational
wave amplitudes and therefore ensures detectability out to larger distances. 
For more details we refer to the recent calculations of Shibata et al. (2005).

\subsection{Neutron star black hole}
The accretion and merger dynamics of neutron star black hole systems is,
mainly due to the larger possible mass ratios, substantially more complicated. 
It is a result of the interplay between gravitational wave emission, mass
transfer and the nuclear EOS  (Rosswog et al. 2004). Gravitational wave
emission will drive the binary towards coalescence, the mass transfer from the
lighter neutron star to the heavier black hole has the opposite tendency.
The influence of the EOS is
twofold. On the one hand it determines the radius of the star, $R_{\rm ns}$,
and therefore the tidal radius, at which the neutron star will be disrupted. 
A simple estimate for this radius is given by
\be
R_{\rm tid}= \left(\frac{{M}_{\rm bh}}{M_{\rm ns}}\right)^{1/3} R_{\rm ns}.
\ee
On the other hand the EOS is responsible for the way the star reacts to 
mass loss, i.e. whether it expands or contracts. For a polytropic star 
the mass-radius relationship is
$R \propto M^\frac{\Gamma-2}{3\Gamma-4}$, where $\Gamma$ is the polytropic 
exponent of the EOS (e.g. Kippenhahn and Weigert 1990). The exponent of the 
mass-radius relation of a polytropic star, $(\Gamma-2)/(3\Gamma-4)$,
is plotted in Fig.~\ref{fig:m-r-expon}. For $\Gamma<2$ the star expands
on mass loss and for $\Gamma>2$ it shrinks on mass loss. For a realistic EOS,
however, the situation is complicated by the fact that $\Gamma$ changes with
density. For the EOS we use (Shen et al. 1998), the neutron star will shrink 
on mass loss for masses above 0.4 \msun and expand otherwise.\\ 
To build up an accretion disk a large ratio of tidal radius to innermost stable
circular orbit, $R_{\rm isco}$,
\be
\frac{R_{\rm tid}}{R_{\rm isco}}= \frac{c^2}{6 G} \frac{R_{\rm ns}}{(M_{\rm
    ns} M_{\rm bh}^2)^{1/3}}, 
\ee
is favorable. Thus, this simple estimate suggests that low masses for 
both the black hole and the neutron star are favorable.\\  
%Unfortunately these interesting systems are technically the most challenging
%as general relativistic effects cannot be calculated using a fixed metric or
%simple approximations to it such as the \Pacz-Wiita potential. Here, the
%final answer can only come from fully general relativistic
%calculations. Lacking the corresponding tools, hydrodynamic simulations have
%so far resorted to purely Newtonian gravity for low black hole masses (Lee
%2000 and 2001, Janka et al. 1999, Rosswog et al. 2004) or approximations to
%general relativity for the higher mass case (Rosswog 2005, Faber et
%al. 2005). Taniguchi et al. recently constructed general relativistic
%quasi-equilibrium sequences for black holes that are much more massive than
%the neutron star. L\"offler et al. (2006) simulated a relativistic head-on
%collision between a neutron star and a black hole.\\ 
Taking together the calculations described in Rosswog (2005) and those of 
Table \ref{runs} the  whole plausible black hole mass range from 3 to 20 \msun 
(Fryer and Kalogera 2001) has been covered. For black holes
below 10 \msun Newtonian potentials and for more massive BHs \Pacz-Wiita
(PW) pseudo-potentials have been used. To mimic the presence of a last stable
orbit in the Newtonian calculations, we set up an absorbing boundary at
$R_{\rm isco}= 6 GM_{\rm BH}/c^2$. For the PW-potential matter is absorbed at
$3 GM_{\rm BH}/c^2$, for details see appendix in Rosswog (2005), but this
material has passed the last stable orbit already and falls 
freely towards the hole, so the exact location of the boundary is
unimportant. In all calculations a non-spinning neutron star mass with
1.4 \msun was used.\\
For illustration, we want discuss two of the runs in more detail: 
A run with a 3 \msun black hole (run NA in Tab.~\ref{runs}) as an 
example for a long-lived, episodic mass transfer, and a 14 \msun run in 
which the neutron star is disrupted at during the first encounter.

%Episodic mass transfer
\subsubsection{Episodic mass transfer}
\label{sec:episodic}
\begin{figure*}
\center{\includegraphics[angle=-90,width=1.0\columnwidth]{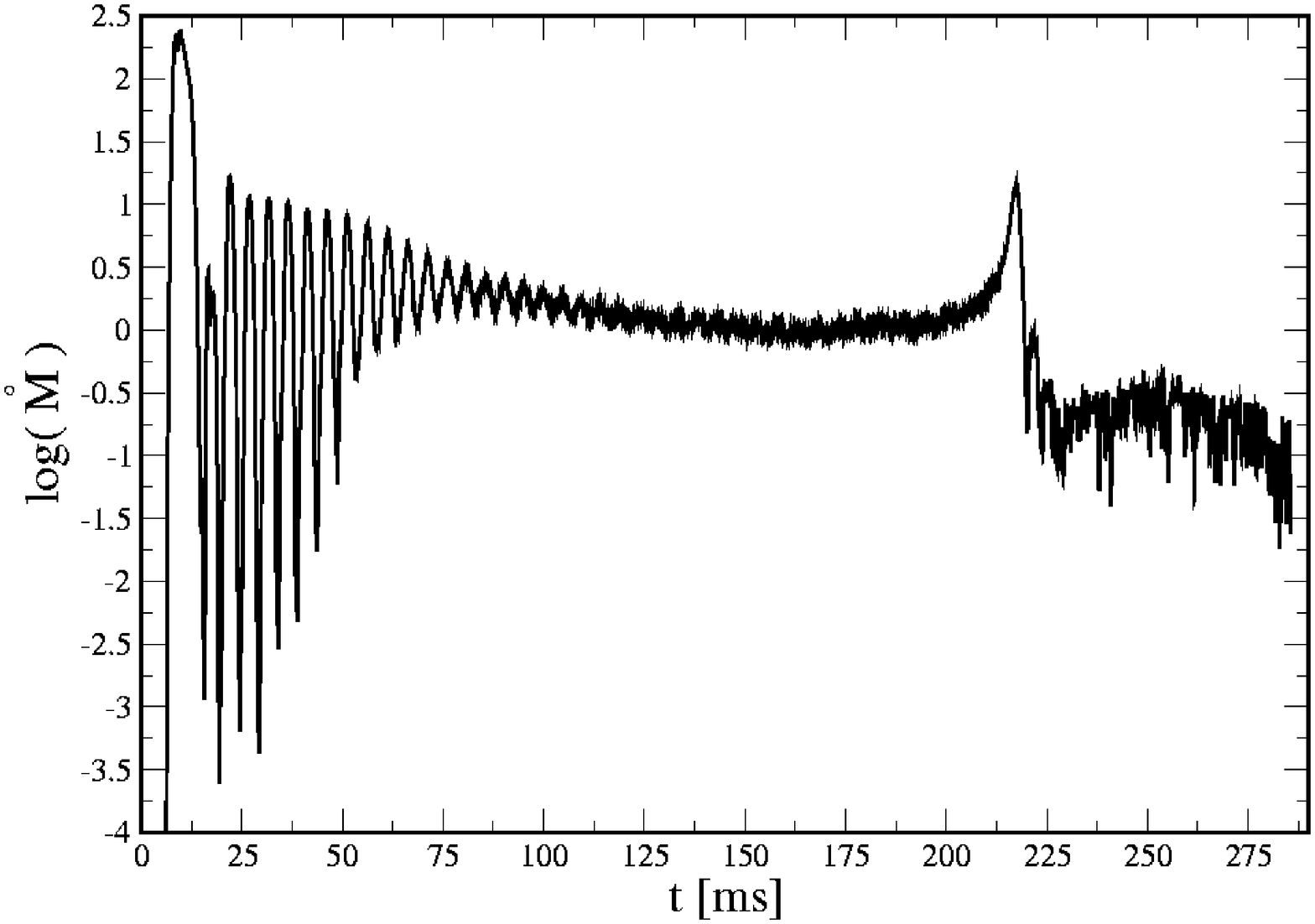}
        \includegraphics[angle=-90,width=1.0\columnwidth]{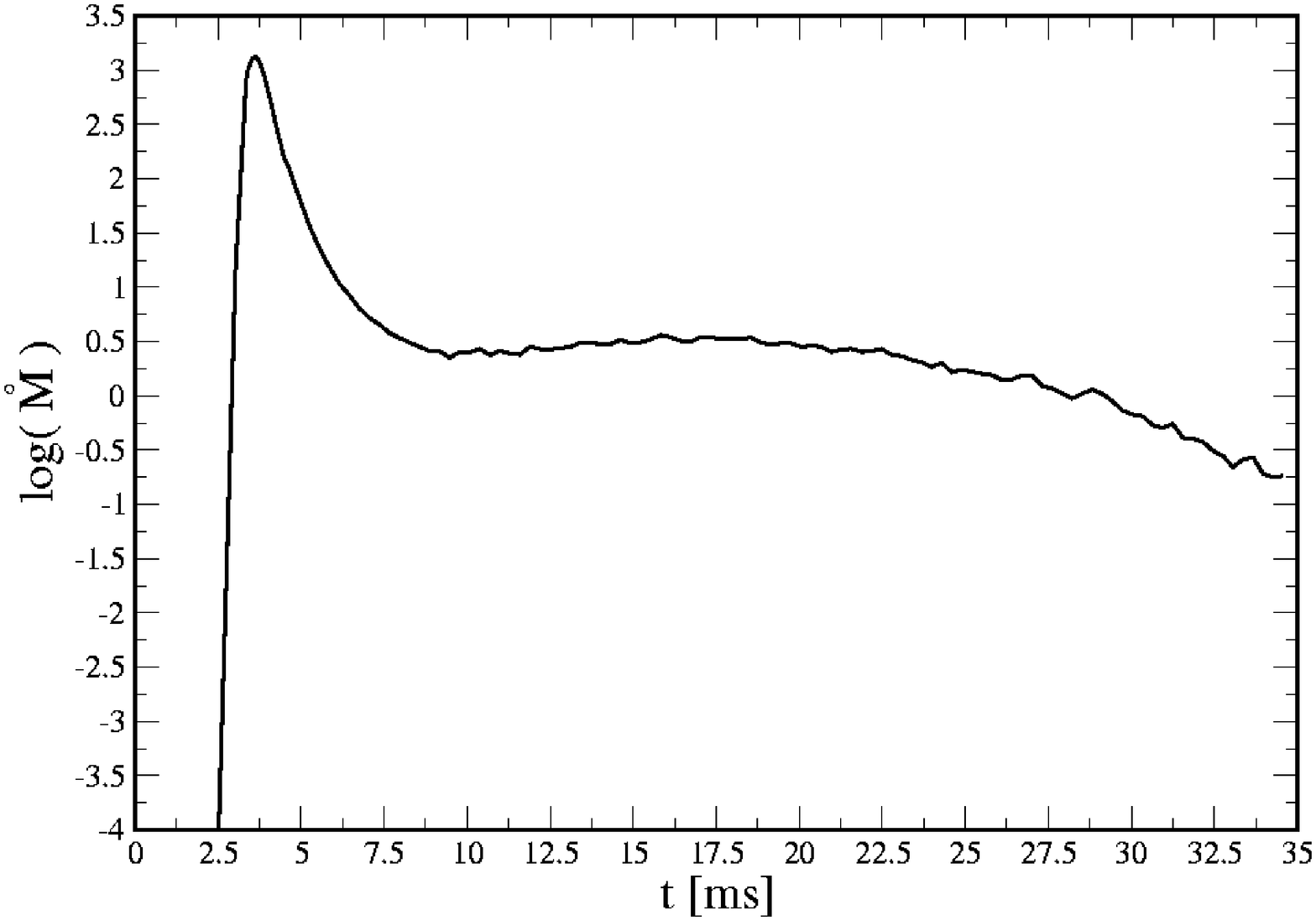}}
  \caption{Left panel: mass transfer (in \Msun/s) into the black hole for 
    the 1.4 \msun and 3 \msun system with episodic mass transfer (run NA in Tab.~\ref{runs}). 
    Right panel: dito  for the 1.4 \msun and 14 \msun system in which the
    neutron star is directly disrupted.}   
  \label{fig:mdot}
\end{figure*}

\begin{figure*}
\center{\includegraphics[angle=-90,width=0.9\columnwidth]{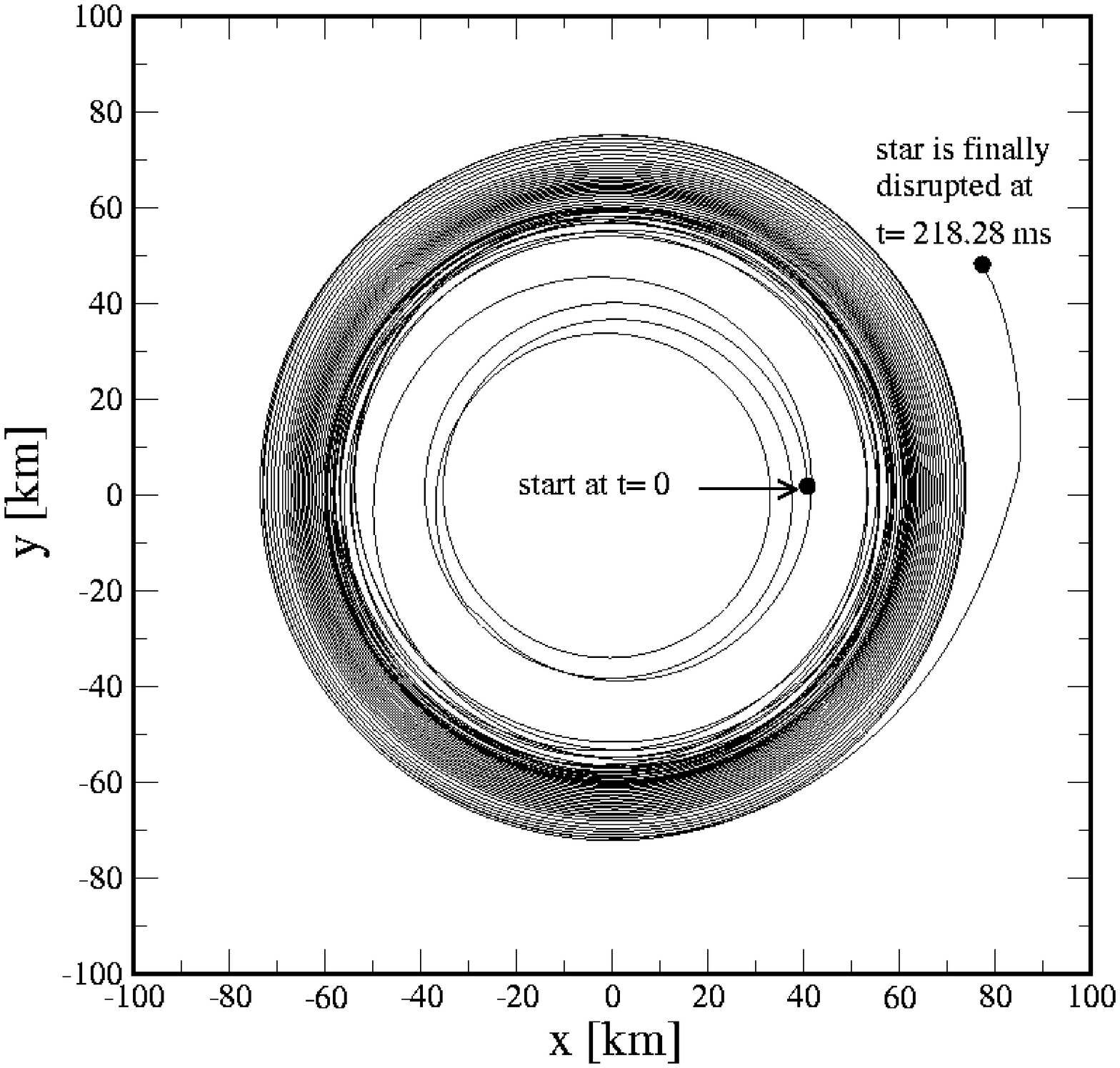}
 \vspace*{-0.5cm}       \includegraphics[angle=-90,width=1.19\columnwidth]{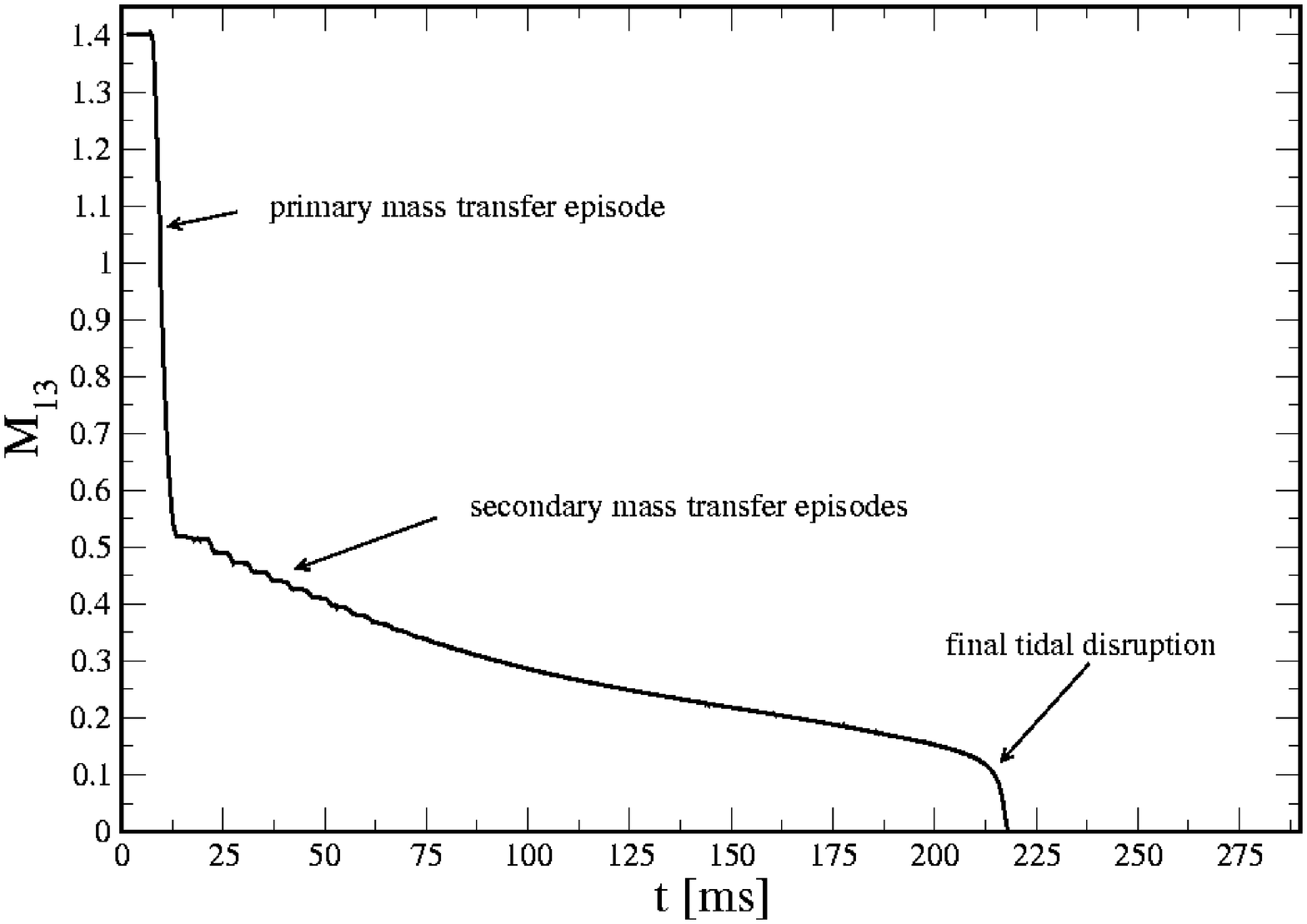}}
  \caption{Left panel: orbit of the mini-neutron star in the neutron star 
                        black hole system (M$_{\rm ns}$= 1.4 \msun, 
                        M$_{\rm bh}$= 3 \Msun) that undergoes episodic
                        mass transfer. After mass transfer 
                        has set in, the neutron star orbits the black hole 
                        for 47 orbital revolutions before it is finally 
                        disrupted.
           Right panel: elvolution of the neutron star mass (defined as the
                        mass with $\rho > 10^{13}$ \gcc) of the same system.}   
  \label{fig:episodic_orbit}
\end{figure*}

In this calculation with a 3 \msun black hole (run NA in Tab.~\ref{runs}) the
neutron star undergoes a long-lived mass transfer without being disrupted, see
Fig.~\ref{fig:nsbh_dynamics}, left column. After a {\em primary mass transfer
  episode} with peak mass transfer rates beyond 100 \Msun/s, see
Fig.~\ref{fig:mdot}, left panel, the neutron star recedes again and quenches
the mass transfer. During this episode its mass is reduced to about 0.5 \Msun,
see right panel of Fig.~\ref{fig:episodic_orbit}.
Subsequently, it undergoes several more mass transfer episodes, after each
episode mass transfer practically shuts off, leaving the neutron star constant
for a short time (see ``step-like'' behaviour in the right panel of
Fig.~\ref{fig:episodic_orbit}). After about 25 episodes the mass transfer 
settles into a stationary state in which the neutron star transfers 
approximately 1 \Msun/s directly into
the hole without forming an accretion disk. Its general tendency during this
episode is to move further away from the black hole, see left panel of
Fig.~\ref{fig:episodic_orbit}. Some material is tidally shed from the
neutron star into a dilute disk around the still orbiting binary. It is only
after 47 orbital periods, at t$\approx$ 218 ms, that the orbiting mini-neutron 
star becomes tidally disrupted and forms an accretion disk around the hole, see
panels two and three in Fig.~\ref{fig:nsbh_dynamics_cont}. This final tidal
disruption sets in when the mini-neutron star reaches its lower mass limit of
about 0.18 \Msun, see right panel of Fig.~\ref{fig:episodic_orbit} (for the mass radius
relationship of the Shen et al. EOS see Fig. 1 in Rosswog et al. (2004)).\\
This long-lived binary phase is imprinted on the gravitational wave signal, see
Fig.~\ref{fig:GW_comparison}, left panel. The signal ``chirps'' until mass
transfer sets in. The primary mass transfer episode reduces the gravitational
wave amplitude by more than a factor of two. While the neutron star is
constantly stripped by the black hole, the gravitational wave amplitude
decreases until the minimum neutron star mass is reached. During this phase
the binary emits at a frequency of $\approx 500$ Hz, close to LIGO's 
sensitivity peak. This phase should be detectable out to a distance of
$d= 10 \; {\rm Mpc} \; (10^{-21}/h_{\rm min})$. Once this stage is
reached, the gravitational wave signal shuts off abruptly as the matter 
settles quickly into an axisymmetric disk. During this final disruption, 
the mass transfer rate increases for a last time beyond 10 \Msun/s, see the
last pronounced peak around 200 ms in the left panel of Fig.~\ref{fig:mdot}.

\subsubsection{Direct disruption} 
\label{sec:direct}

For more massive black holes, the self-gravity of the neutron star can be
overcome already during the first, primary mass transfer. An example of such a
case with a 14 \msun black hole is shown in Fig.~\ref{fig:nsbh_dynamics},
right column.
Although the neutron star is disrupted, a density peak remains visible in the
debris spiral arm until the end of the simulation. In this case an 
accretion disk forms, but
large parts of this disk are inside the innermost stable circular orbit of the
black hole and therefore falling with large radial velocities into the
hole. This disks are geometrically thin and, apart from a spiral shock from the
self-interaction of the accreted matter, essentially cold. A detailed
discussion of this calculation can be found in Rosswog (2005). During the first
approach mass is transferred into the black hole at a peak rate of more than 
1000 \Msun/s, see right panel of Fig.~\ref{fig:mdot}.\\
Again the dynamics leaves a clear imprint on the gravitational wave signal:
after the chirp stage the signal immediately dies off, see
Fig.~\ref{fig:GW_comparison} right panel. At peak amplitude, the
signal should be visible out to 
$d= 250 \; {\rm Mpc} \; (10^{-21}/h_{\rm min})$.\\ 
For the higher mass black holes, we do not find accretion disks that are
promising to produce energetic GRBs. While some systems do form low-mass 
disks, see Fig.~\ref{fig:nsbh_dynamics}, right column, for a 14 \msun BH, 
for black holes with $M\ge 18$ \Msun, nearly the whole neutron
star is directly fed into the hole, only a small fraction ($<$ 0.08 \Msun)
receives enough angular momentum to be ejected in form of a half-ring (see
Figure 3 in Rosswog 2005) of neutron-rich debris material.\\

\subsection{Constraints on the nuclear equation of state from gravitational
               waves} 

\begin{figure*}
\center{\includegraphics[angle=-90,width=1.7\columnwidth]{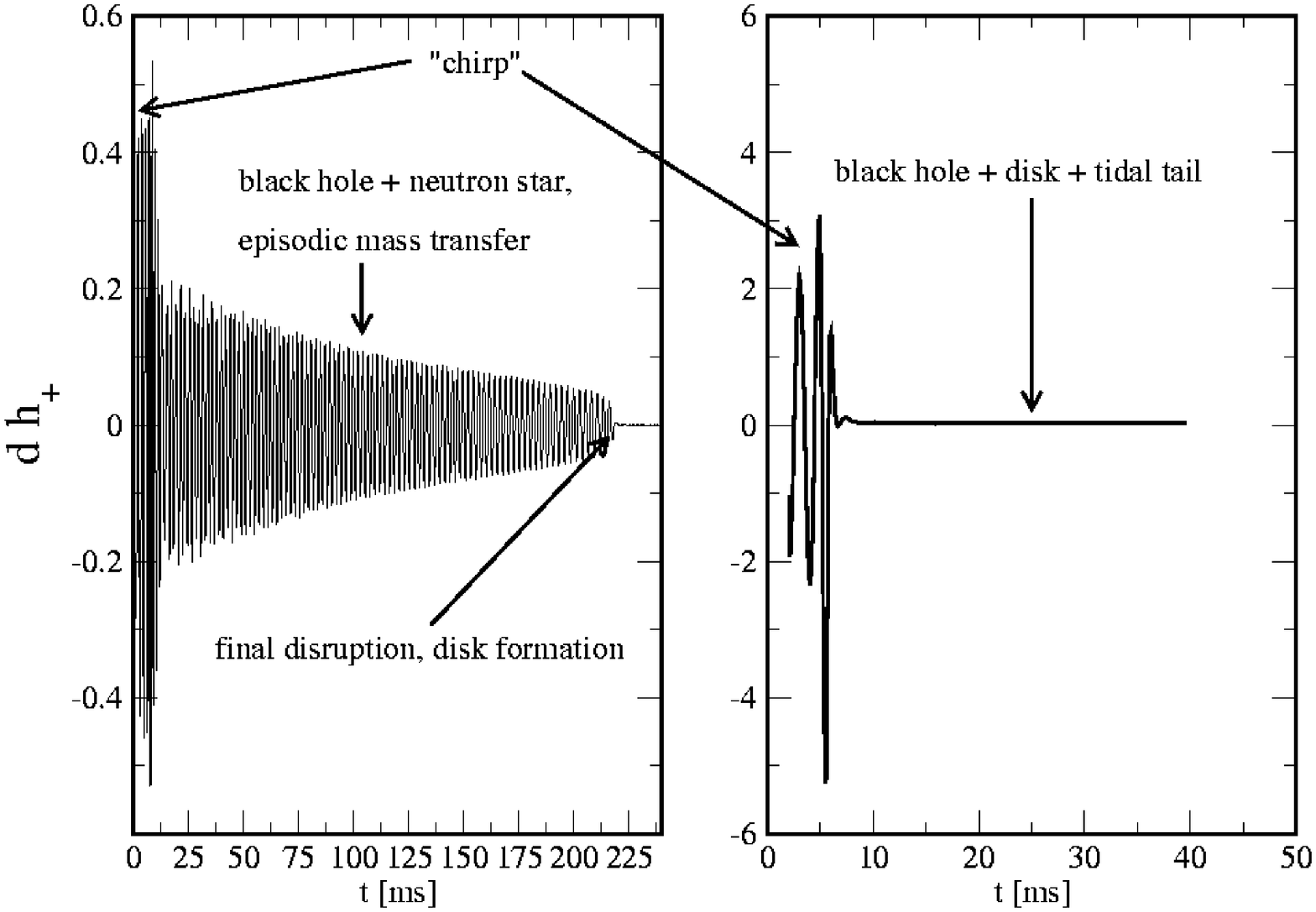}
}
  \caption{Left panel: gravitational wave signal of a binary system that
    undergoes episodic mass transfer (M$_{\rm ns}$= 1.4 \Msun, M$_{\rm bh}$= 3 \Msun).
    Right panel: gravitational waves for the 1.4 \msun and 14 \msun system in which the
    neutron star is directly disrupted (run II of Rosswog (2005)).}   
  \label{fig:GW_comparison}
\end{figure*}

As mentioned earlier, the neutron star equation of state (EOS) could be 
seriously constrained by a detected gravitational wave signal. We want to
briefly summarize gravitational wave signatures that carry information on the
nuclear EOS.
\bi
\item A simple estimate for the neutron star radius can be obtained using
  Newtonian gravity and $\nu_{\rm GW}= 2 \nu_{\rm orb}$, where $\nu_{\rm GW}$
  and $\nu_{\rm orb}$ are the gravitational wave and orbital frequencies.  If
  one assumes an equal mass binary, that the peak occurs once the neutron
  star surfaces touch and that changes due to tides are negligible, one finds
\begin{eqnarray}
R_{\rm ns}&=& \left( \frac{G M}{8 \pi^2 \nu_{\rm p}^2} \right)^{1/3}\nonumber\\
          &=& \hspace*{-0.3cm}16.8 \; {\rm km} \; \left(\frac{M}{2.8 
{\rm M}_\odot}\right)^{\frac{1}{3}} 
  \left(\frac{1 {\rm MHz}}{\nu_{\rm p}}\right)^{\frac{2}{3}}. 
\end{eqnarray}
Here, $\nu_{\rm p}$ is the peak GW-frequency at the end of the chirp phase
and $M$ the total mass of the binary.\\
More accurately, using quasi-equilibrium sequences of neutron star binaries,
  the compactness $GM/Rc^2$ can be determined from the final deviation of the
  gravitational wave energy spectrum from a point mass binary signal (Faber et
  al. 2002). If additionally the neutron star masses are known from the
  inspiral signal (Cutler and Flanagan 1994), the neutron star radius can be
  derived and severe constraints on the equation of state can be imposed.
\item The stiffness of the EOS influences the post-merger behaviour of the
  remnant. If the EOS is stiff enough, the central object will keep a triaxial,
  ellipsoidal shape for many orbital periods and thus will continue to emit 
  gravitational waves (e.g. Rasio and Shapiro 1994), 
  before it (probably) finally  collapses into a black hole. For a soft EOS it
  will quickly become axi-symmetric and the GW-signal will die off fast.\\
  The influence of quark matter on the gravitational wave
  signal has recently been investigated by Oechslin et al. (2004).
\item As was shown in Sec.\ref{sec:episodic}, NSBH systems with low black 
  hole masses can undergo a
  long-lived (in comparison to orbital time), mostly episodic mass transfer
  activity. Test calculations with a soft ($\Gamma=2$) polytrope (Rosswog et
  al. 2004) indicate that this behaviour is specific for stiff equations of
  state. Moreover, from a gravitational wave signal such as the one shown
  Fig.~\ref{fig:GW_comparison}, left panel, the time when the neutron star
  reaches its minimum mass can be directly read off: at this point the 
  gravitational wave signal suddenly shuts off. 
\ei

\section{Accretion disks}
\label{sec:accretion}
The accretion disks produced in a compact binary merger are in several 
respects
different from a standard Shakura-Sunyaev (1973) accretion disk. Having just
been built up in a violent disruption, they are far from a steady state and
completely opaque to photons. Cooling is only possible via neutrino
emission. The corresponding processes, however, are not generally fast enough
to cool the disks on a timescale comparable to the dynamical timescales. The
disks often span the full range from neutrino opaque to completely transparent
with relatively large semi-transparent transition regions making an analytical
treatment without restrictive assumptions difficult. Over the last years
several groups have worked in different approximations on the neutrino-cooled
disk problem (e.g. Popham et al. 1999, Narayan et al. 2001, Kohri and
Mineshige 2002, DiMatteo et al. 2002). Recently, equilibrium disks around
rotating black holes have been constructed by Chen and Beloborodov (2006).
Neutrino opacity effects have been included in several of the recent 
numerical simulations (Ruffert et al. 1997, Rosswog and Liebend\"orfer 2003, 
Lee et al. 2005).
%DNS
\subsection{Double neutron star mergers}
\begin{figure}[!t]
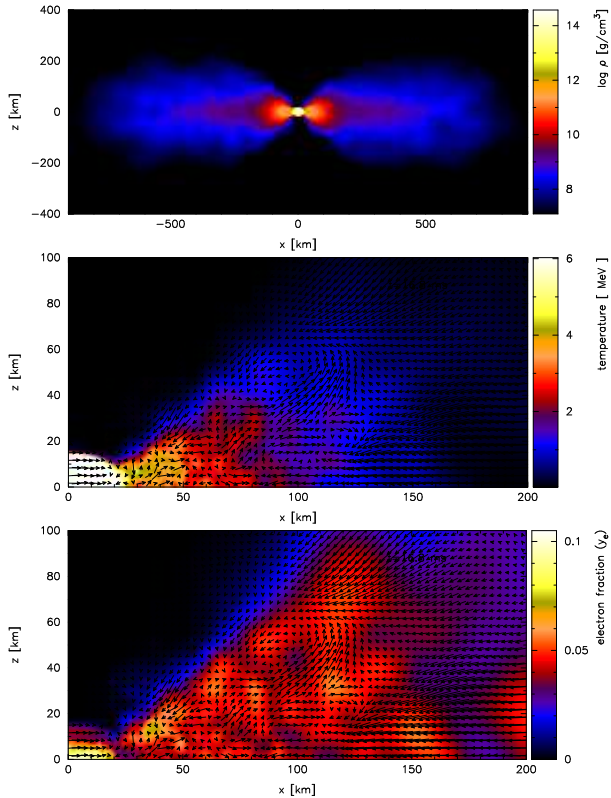

%  \includegraphics[angle=-90,width= \columnwidth]{disk_dens_global_fd135.ps}
%  \includegraphics[angle=-90,width= \columnwidth]{disk_T_velocity_fd135_zoom1.ps}
%  \includegraphics[angle=-90,width=
%  \columnwidth]{disk_Ye_velocity_fd135_zoom1.ps}
%
%
  \includegraphics[angle=-90,width= \columnwidth]{disk_dens_global_fd135.ps}
  \includegraphics[angle=-90,width= \columnwidth]{disk_T_velocity_fd135_zoom1.ps}
  \includegraphics[angle=-90,width= \columnwidth]{disk_Ye_velocity_fd135_zoom1.ps}
  \caption{Vertical structure of the debris disk resulting from an 
   equal mass (1.4 \Msun) double neutron star merger (run DA, see Table 1, 
  at t= 16.88 ms). 
  Panel one shows the global mass density distribution in the XZ-plane; 
  panel two zooms into inner disk region, color-coded is the temperature. To
  enhance the temperature contrasts in the disk the upper limit of the
  colorbar has been fixed to 6 MeV.
  Panel three shows the distribution of the electron fraction, $Y_e$; in
  panels two and three the velocity field is overlaid.}
  \label{fig:DNS_disk}
\end{figure}

We show in Fig.~\ref{fig:DNS_disk} disk properties of run DA that we consider
representative for the DNS case. The dominant neutrino-emitting parts of the 
disks have
densities in the range from $\sim 10^{13}$ down to $\sim 10^{10}$ \gcc, see
panel one, the temperatures in these regions are $\sim 3$ MeV. 
At the time of the merger, the neutron stars are still very close to cold
$\beta$-equilibrium as the tidal interaction is too short to 
cause substantial compositional changes at the prevailing temperatures
($T<10^8$ K; Lai 1995). The weak interaction rates during the merger are 
sufficiently slow that the matter in the remnant is at $\sim$ 15 ms 
still close to its initial electron fraction, $Y_e\sim 0.05$, see panel 
three in Fig.~\ref{fig:DNS_disk}. Therefore the assumption of
$\beta$-equilibrium is not justified.\\
The central object is apart from its surface layers completely opaque to the
neutrinos. The inner parts of the debris torus still have large optical
depths, see Fig. 8-10 in Rosswog and Liebend\"orfer (2003), only the outer
parts are neutrino transparent. As the disks cannot cool on a dynamical 
timescale they are puffed-up and advection dominated. Inflow towards the 
central object proceeds mainly along the equatorial plane (see arrows 
in panel two and three) and along the disk surface, the flow inside the 
disks shows convective circulation. For a further discussion of the
circulation patterns see Lee and Ramirez-Ruiz (2002), 
Rosswog and Davies (2002) and Lee et al. (2005).\\  
%NSBH
\subsection{Neutron star black hole mergers}
In the neutron star black hole case we find it much harder to build up massive
accretion disks. In the low-mass cases the episodic mass transfer prevents 
disk formation until the neutron star is finally disrupted once it has
been stripped down to its minimum mass of about 0.18 \msun. Such a case was 
shown in detail in Figs.~\ref{fig:nsbh_dynamics} and 
\ref{fig:nsbh_dynamics_cont}. The resulting disk as seen in panel six of
Fig.~\ref{fig:nsbh_dynamics_cont} has about 0.05 \Msun. For the cases with
larger black hole mass, the neutron star is completely disrupted early on, see 
Sec.~\ref{sec:direct}, but close to the innermost stable circular orbit.
As a consequence, matter falls with large radial velocities into the hole 
within about one orbital period. These disks are essentially cold (apart 
from a spiral shock that occurs due to self-interaction of the accretion 
stream, see Fig. 6 in Rosswog 2005), low in density and, contrary to the 
neutron star merger disks, they are geometrically thin.
For even higher black hole masses the neutron star is nearly swallowed
completely without any disk formation. For a further discussion we refer to
Rosswog (2005).\\
These result seems to depend on the nuclear equation of state. 
Janka et al. (1999) who use the softer Lattimer-Swesty EOS
find disks that are more promising for GRBs. \\

\begin{figure*}
\center{\includegraphics[angle=-90,width=1.03\columnwidth]{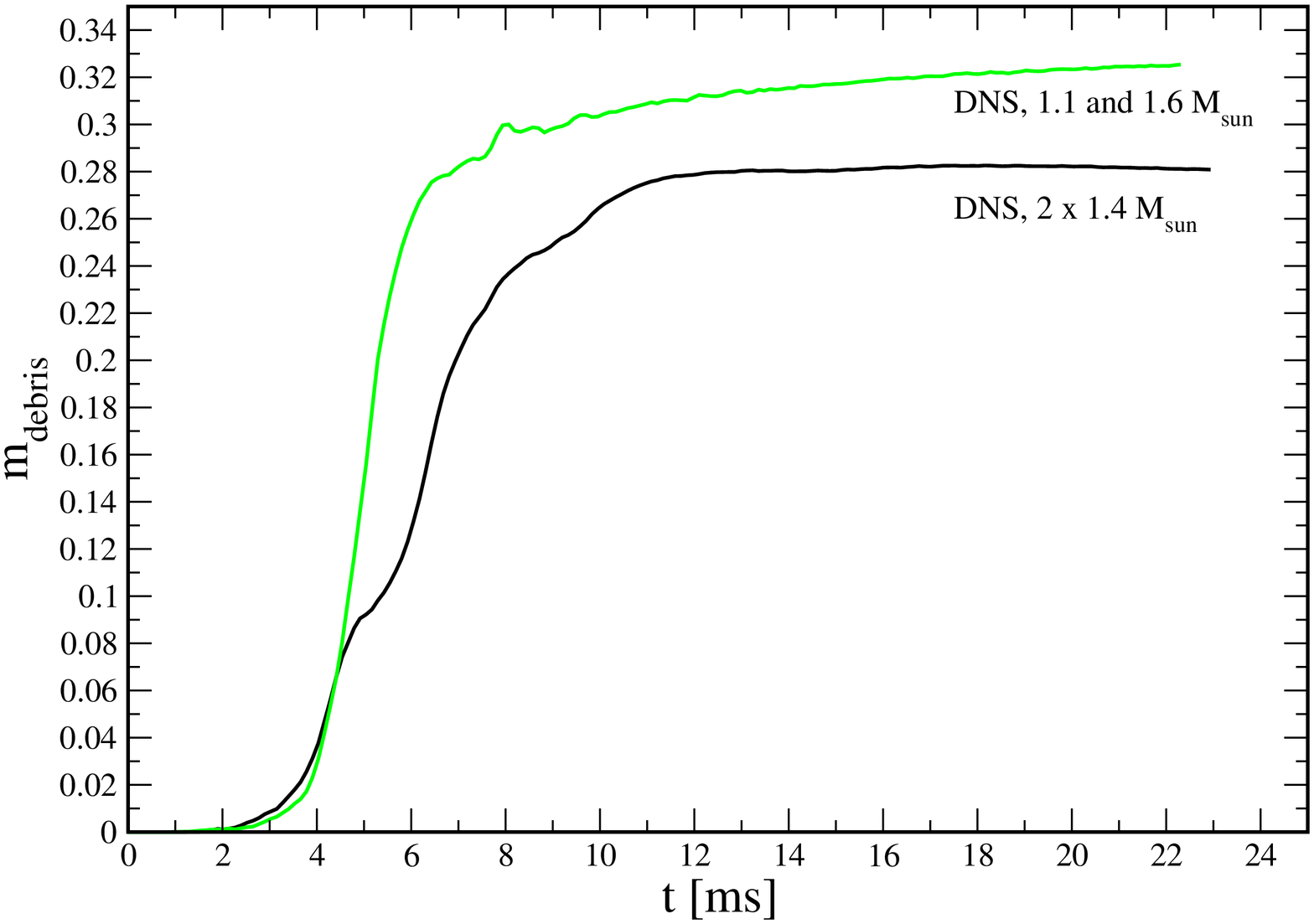}
        \includegraphics[angle=-90,width=1.\columnwidth]{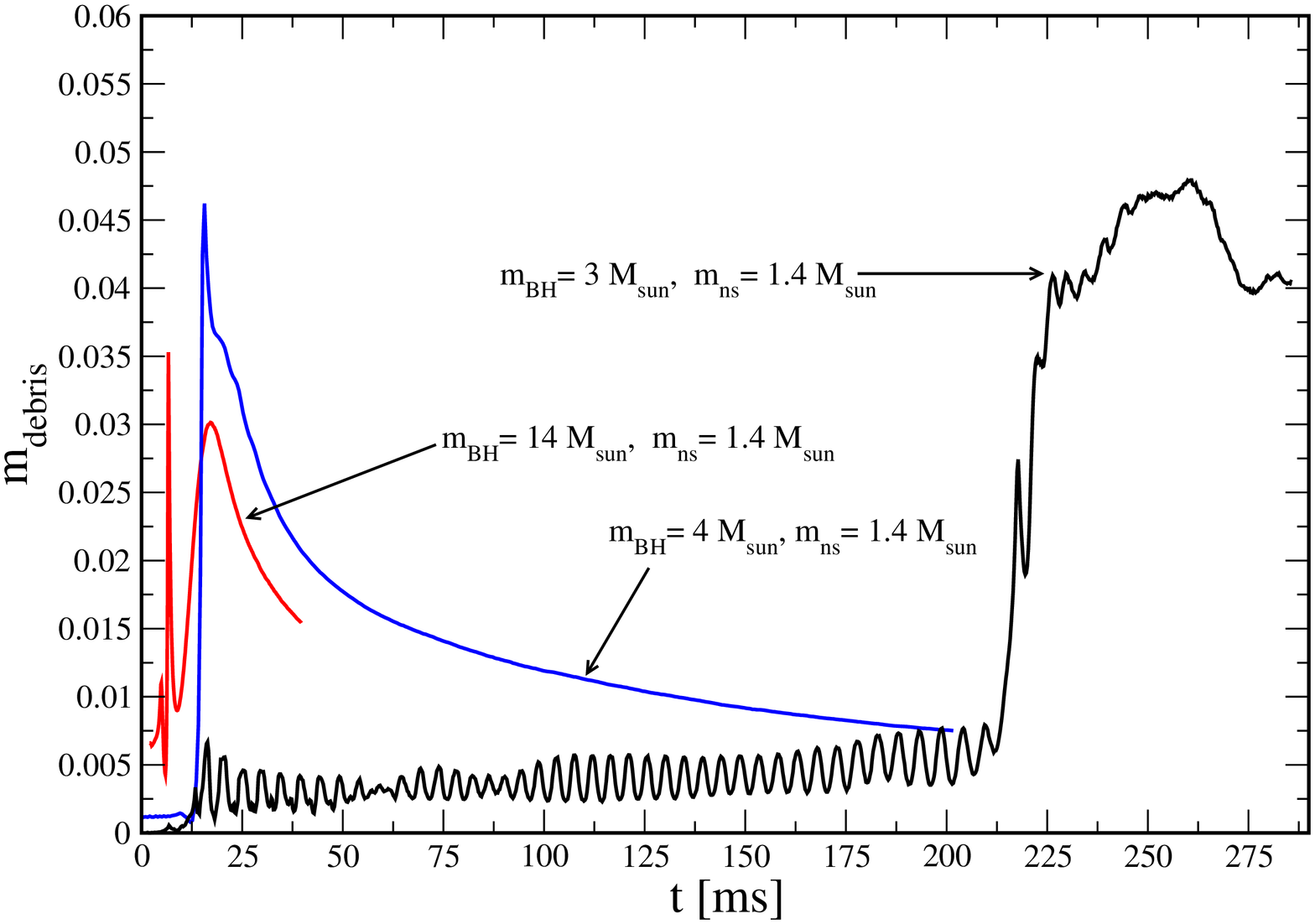}}
  \caption{Comparison between double neutron star and neutron star black hole mergers.
        Left panel: debris mass (gravitationally bound, 
        $\rho < 10^{13}$ \gcc) for the double neutron star case. 
        Right panel: debris mass for the neutron star black hole cases. Note
        the different scales on the axes of both panels.}   
  \label{fig:bound_debris}
\end{figure*}

\begin{figure*}
  \centerline{\includegraphics[angle=-90,width=1.1\columnwidth]{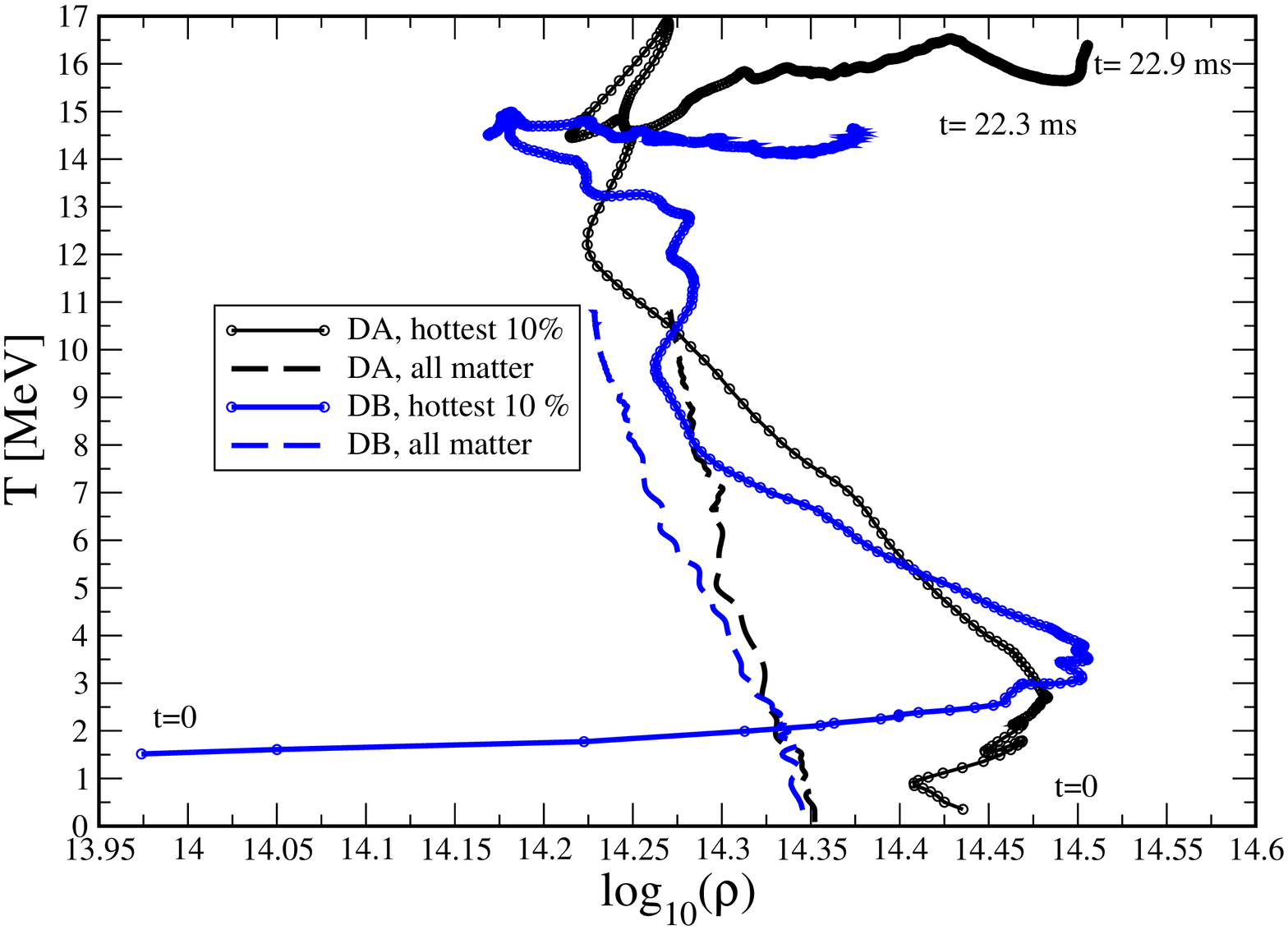}
\includegraphics[angle=-90,width=1.1\columnwidth]{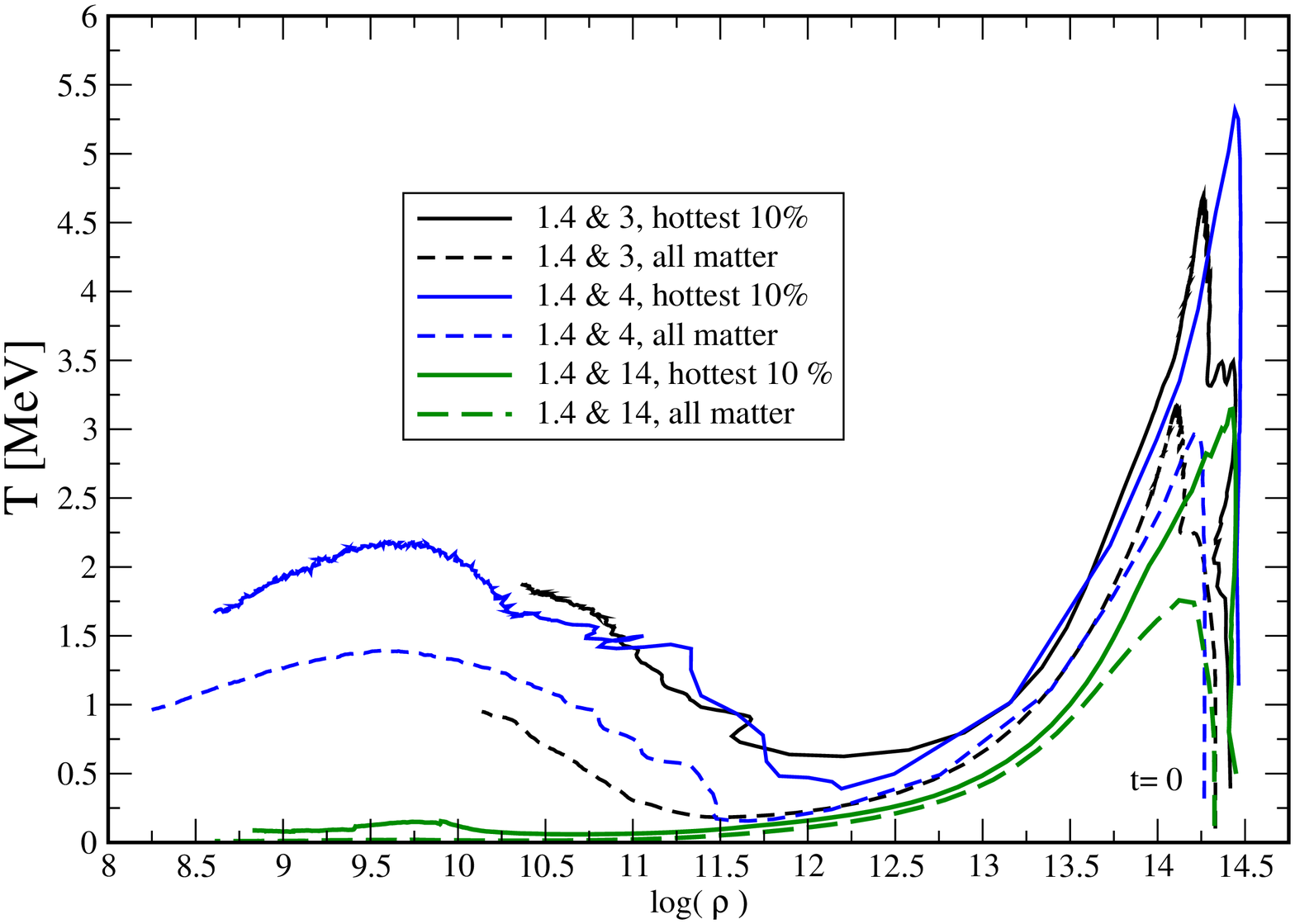}}
  \caption{Comparison between double neutron star and neutron star black hole
    mergers. Left panel: Trajectories in the $\rho-T$-plane for
     double neutron star cases (run DA and DB), right panel: typical neutron
 star black hole cases. For each case, the trajectories of the hottest 10\% of
    the material and an average over all matter are shown.}
  \label{fig:DNS_rho_T_trajecories}
\end{figure*}

To quantify the amount of present debris material for both the DNS and the
NSBH cases, we use the mass that is gravitationally bound and has a 
density $\rho<10^{13}$ \gcc.
The evolution of this debris mass is shown in Fig.~\ref{fig:bound_debris}. 
In the DNS cases the debris mass is around 0.3 \Msun, for the black hole cases
it is about one order of magnitude smaller, at maximum $\approx 0.05$ \Msun. 
In run NA, which is described in detail in Sec.~\ref{sec:episodic}, the mass 
transfer is constantly driving neutron star oscillations, see 
Fig.~\ref{fig:bound_debris}, right panel. The debris mass only increases
beyond 0.01 \msun after the neutron star is completely disrupted at t$\approx
218$ ms.\\
To further illustrate the different thermodynamic condition prevailing in the
debris of DNS and NSBH remnants, we show in 
Fig.~\ref{fig:DNS_rho_T_trajecories}
trajectories in the $\rho-T$-plane. The left panel refers to the two
neutron star cases, the right panel shows typical trajectories of NSBH
systems. The solid lines always refer to the average of the hottest 
10 \% of the debris, the dashed lines show the average conditions of all
matter. While in the neutron star case the hottest material is always at very
high densities ($>10^{14}$ \gcc), the hottest regions in the NSBH cases 
move during the evolution down to densities of $\le 10^{10}$\gcc and below. In
the direct disruption case that was discussed in Sec.~\ref{sec:direct}, 
the average temperature of even the hottest 10 \% of the material quickly drops
below 1 MeV (solid green curve).


\section{Neutrino emission}
\label{sec:neutrinos}

%DNS
\subsection{Double neutron stars}
\begin{figure}
\center{\includegraphics[angle=-90,width=1.1\columnwidth]{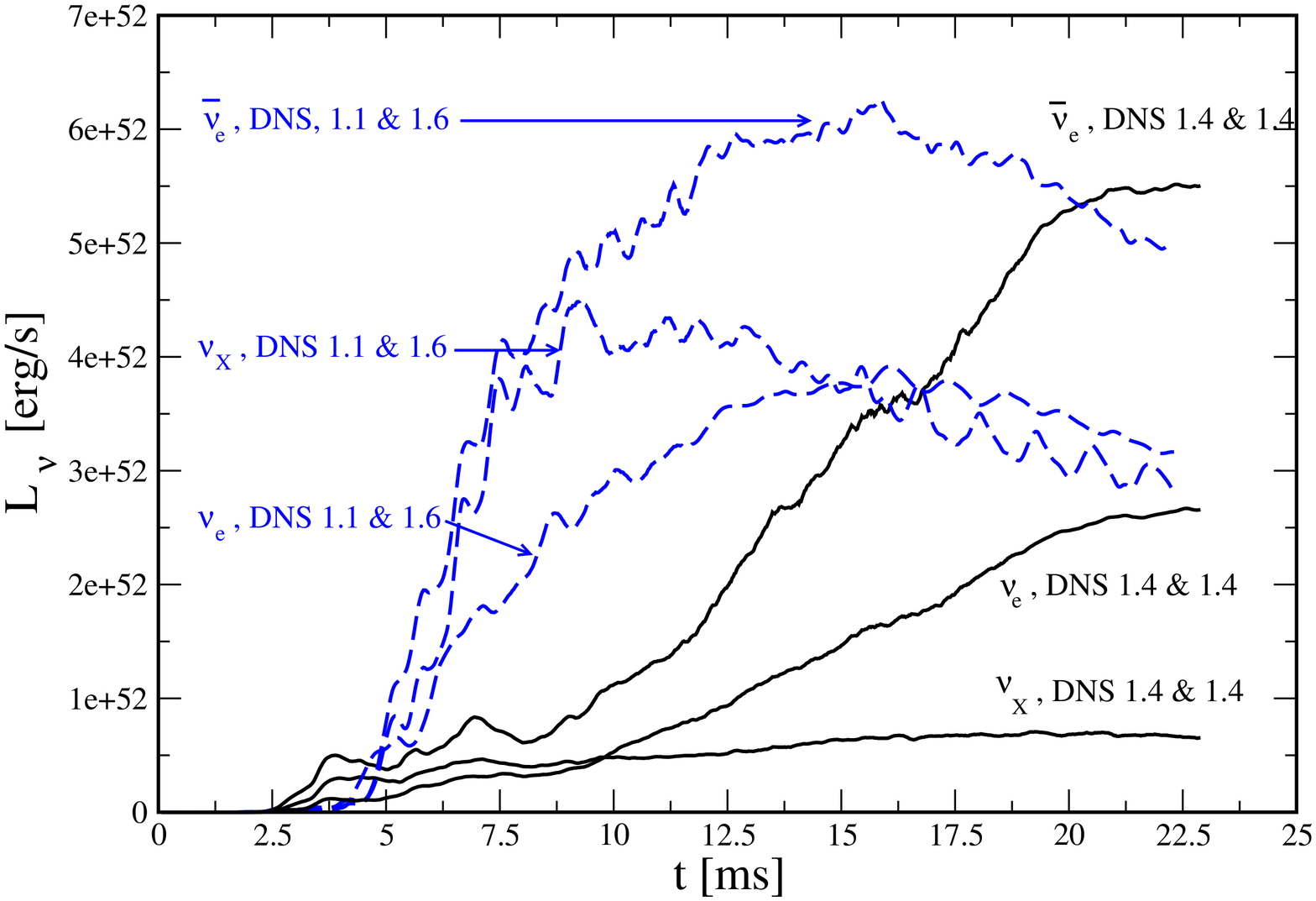}}
  \caption{Neutrino luminosities of the double neutron star merger
    cases. Solid lines refer to the case with 2 x 1.4 \Msun, the dashed lines 
to the 1.1 and 1.6 \Msun case.}   
  \label{fig:nu_lum_DNS}
\end{figure}

The debris in the DNS case is very neutron-rich and hot, so the
neutrino luminosities are generally dominated by electron anti-neutrinos
from positron captures onto free neutrons, $e^+ + n \rightarrow p +
\nu_e$. Electron neutrinos, mainly from electron captures,  
$e + p \rightarrow n + \bar{\nu}_e$, are second most important and 
followed by the heavy lepton neutrinos 
$\nu_\mu,\bar{\nu}_\mu,\nu_\tau,\bar{\nu}_\tau$ that we generally refer to as
$\nu_x$. While the $\nu_e$ and $\bar{\nu}_e$ are emitted predominantly from
the inner disk regions or the surface layers of the central object, the heavy
lepton neutrinos $\nu_x$ are mainly produced in the hot high-density regions
inside the central object. Since the heavy lepton neutrinos are not 
absorbed by nucleons, the surrounding matter is more transparent to them. 
Therefore, they can escape from the hotter, high density regions more easily,
which is why they have the largest average energy. The average energies of 
the neutrino species are relatively 
robust against changes in the system parameters, typically $\langle 
E_{\nu_e} \rangle \approx 7$ MeV, $\langle E_{\bar{\nu}_e} \rangle \approx 12$
MeV and $\langle E_{\nu_x} \rangle \approx 23$ MeV. In the unlikely case
that neutrinos from a neutron star merger would be detected, the substantial 
drop in the $\nu_x$, but only a milder drop in the $\nu_e$ and
$\bar{\nu}_e$ luminosity would indicate the moment when the central objects
collapses into a black hole.\\
The neutrino luminosities after a DNS merger increase 
smoothly until a stationary state is reached, see Fig.~\ref{fig:nu_lum_DNS}. 
At this stage the luminosities of the numerical models can be roughly fit by
\be
L_{\nu}^{tot}= L_1 \left(\frac{M}{2.6 \; {\rm M}_\odot}\right)^\alpha
\label{eq:lnu_tot}
\ee
with $L_1= 6.8\cdot10^{52}$ erg/s,  $\alpha=4.5$ and $M$
being the total binary mass. This formula does not apply to cases where an
Algol-like impact occurs (run DB), in such a case the neutrino-luminosities
are higher due to the very strong shock heating at moderate densities.
The contribution of the different neutrino flavors to the total
luminosity changes with the details of the system under consideration, but 
they are approximately given by 
$L_{\nu_e} \approx 0.3 \cdot L_{\nu}^{tot}$, $L_{\bar{\nu}_e}
\approx 0.5 \cdot L_{\nu}^{tot}$ and $L_{\nu_x} \approx 0.2 \cdot
L_{\nu}^{tot}$.\\
The disks that were produced in neutron star black hole calculations were only
of moderate masses and temperatures and therefore produced much lower
luminosities than in the neutron star case. Calculations with the softer EOS
of Lattimer and  Swesty (1991) (Janka et al. 1999), find results not too
different from the DNS case.

\section{GRBs}
\label{sec:GRBs}
Compact binary mergers have been recognized as possible
central engines of GRBs since many years (Blinnikov et al. 1984, 
Eichler et al. 1998, \Pacz 1991, Narayan et al. 1992). For general 
reviews on the manifold aspects of gamma-ray bursts we refer to recent reviews
(e.g. Piran 2005 or \Mesz 2006).\\
To attain the ultra-relativistic motion required to explain both the
short-time variability and the non-thermal spectrum of GRBs, large amounts
of energy have to be deposited in a region of space that is practically devoid
of baryons. Two popular mechanisms to achieve this are
the annihilation of neutrino anti-neutrino pairs (Eichler et al. 1989,
Mochkovitch et al. 1993, Ruffert et al. 1997, Popham et al. 1999, Asano and
Fukuyama 2003, Rosswog and Ramirez-Ruiz 2002, Rosswog et al. 2003) and 
mechanisms that rely on (ultra-)strong magnetic fields (Narayan
et al. 1992, Usov 1992, Duncan and Thompson 1992, Thompson 1994, \Mesz and
Rees 1997, Kluzniak and Ruderman 1998, Lyutikov et al. 2003, Rosswog et
al. 2003).\\ 
%\subsection{Double neutron stars}
\subsection{$\nu_i-\bar{\nu}_i$-annihilation}
The energy deposition rate per volume via neutrino anti-neutrino annihilation
(see Ruffert et al. 1997) can be written is discretised form as

\begin{eqnarray}
&&Q_{\nu \bar{\nu}}(\vec{r})=\sum_{i= e,\mu,\tau} Q_{\nu_i \bar{\nu}_i}(\vec{r})\nonumber\\
&=&\sum_{i= e,\mu,\tau} A_{1,i} 
\sum_k \frac{L^k_{\nu_i}}{d_k^2} \sum_{k'} \frac{L^{k'}_{\bar{\nu}_i}}{d_{k'}^2}
[ \langle E_{\nu_i}\rangle^k +  \langle E_{\bar{\nu}_i}\rangle^{k'}] 
\mu_{kk'}^2 \nonumber\\
 &+&\sum_{i= e,\mu,\tau} A_{2,i} \sum_k \frac{L^k_{\nu_i}}{d_k^2} \sum_{k'} 
\frac{L^{k'}_{\bar{\nu}_i}}{d_{k'}^2}
 \frac{\langle E_{\nu_i}\rangle^k +  \langle E_{\bar{\nu}_i}\rangle^{k'}}
{\langle E_{\nu_i}\rangle^k  \langle E_{\bar{\nu}_i}\rangle^{k'}}
\mu_{kk'}\label{eq:ann}
\end{eqnarray}
where the index $i$ labels the type of neutrino.  Here $L^k$ is the
neutrino luminosity of grid cell $k$, $d_k$ is the distance from the
centre of grid cell $k$ to the point $\vec{r}$, $d_k=
|\vec{r}-\vec{r}_k|$, $\langle E_{\nu_i}\rangle^k$ is the average
neutrino energy in grid cell $k$, $\mu_{kk'}= 1-\cos \theta_{kk'}$ and
$\theta_{kk'}$ is the angle at which neutrinos from cell $k$ encounter
anti-neutrinos from cell $k'$ at the point $\vec{r}$. The constants 
are given by
$A_{1,e}=\frac{1}{12\pi^2}\frac{\sigma_0}{c(m_ec^2)^2}
[(C_V-C_A)^2+(C_V+C_A)^2] $,
$A_{1,\mu}=A_{1,\tau}=\frac{1}{12\pi^2}\frac{\sigma_0}{c(m_ec^2)^2}
[(C_V-C_A)^2+(C_V+C_A-2)^2] $, $A_{2,e}=
\frac{1}{6\pi^2}\frac{\sigma_0}{c} [2 C_V^2-C_A^2]$,
$A_{2,\mu}=A_{2,\tau}= \frac{1}{6\pi^2}\frac{\sigma_0}{c} [2
(C_V-1)^2-(C_A-1)^2] $, where $C_V= 1/2+2 \sin^2 \theta_W$, $C_A=
1/2$, $\sin^2\theta_W= 0.23$ and $\sigma_0= 1.76 \times 10^{-44}$
cm$^2$.

The thick-disk geometry that is a natural outcome of a double neutron star 
merger, see Fig.~\ref{fig:DNS_disk}, is favorable for an efficient annihilation
 as neutrinos have enhanced probability to collide head-on. The most
interesting regions for the energy deposition are the centrifugally
baryon-cleaned regions along the original binary rotation axis.
Contours of the matter density and the annihilation energy deposition rate can
be found in Fig. 2 in Rosswog et al. (2003).
%which fraction of the energy is transformed into pair plasma
The deposition of a large amount of energy in a photon-pair plasma between the 
funnel walls results in a bipolar outflow along the binary rotation axis. 
The ratio of deposited annihilation and rest mass energy determines the 
asymptotic Lorentz-factors of the resulting outflow. For a double neutron star 
merger these Lorentz-factors lie between a few dozens and a few times $10^4$
(see Rosswog et al. 2003, Figs. 4-6). The total energy contained in 
this relativistic outflow is $\sim 10^{48}$ erg 
(Rosswog and Ramirez-Ruiz 2003).\\
\begin{figure*}
\center{\includegraphics[angle=-90,width=1.03\columnwidth]{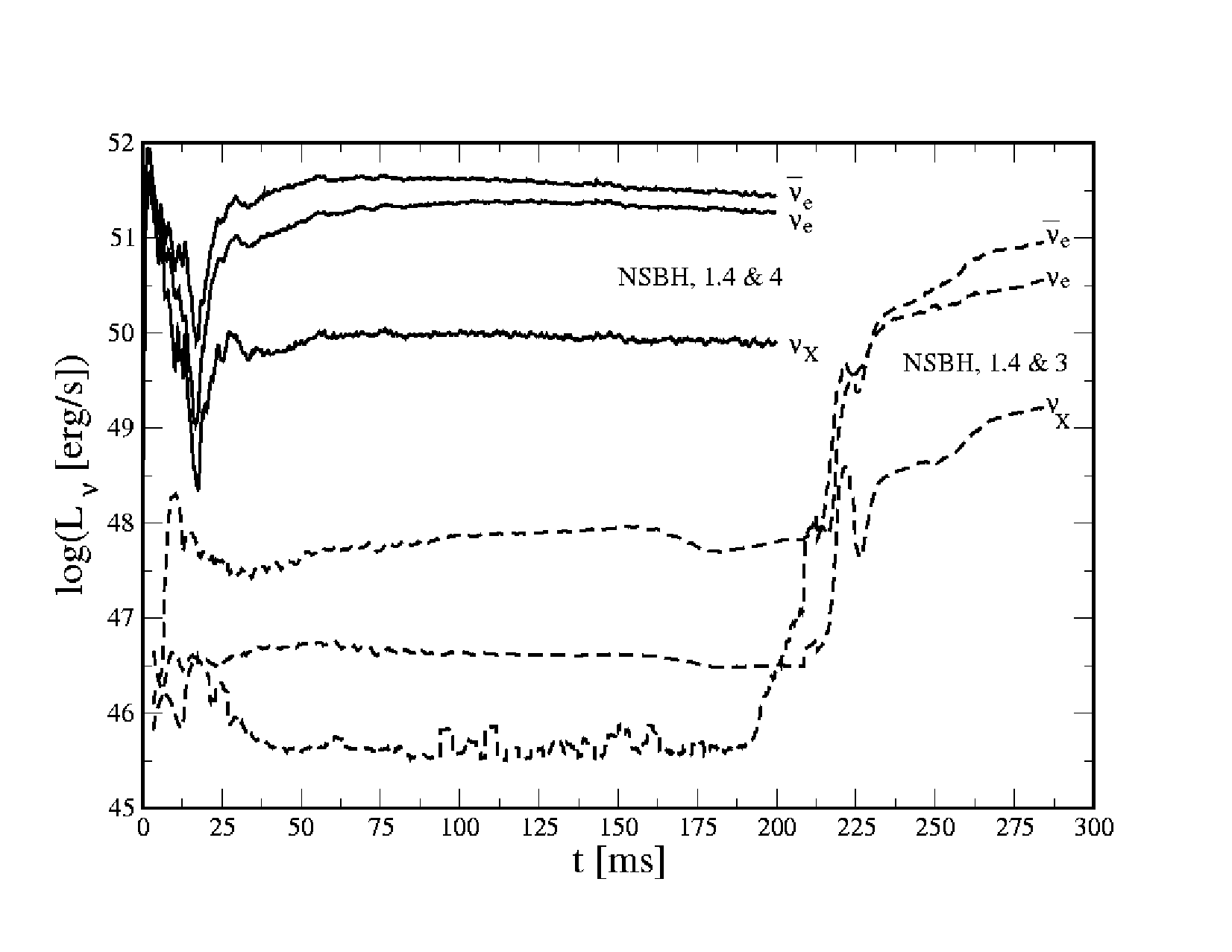}
\includegraphics[angle=-90,width=1.\columnwidth]{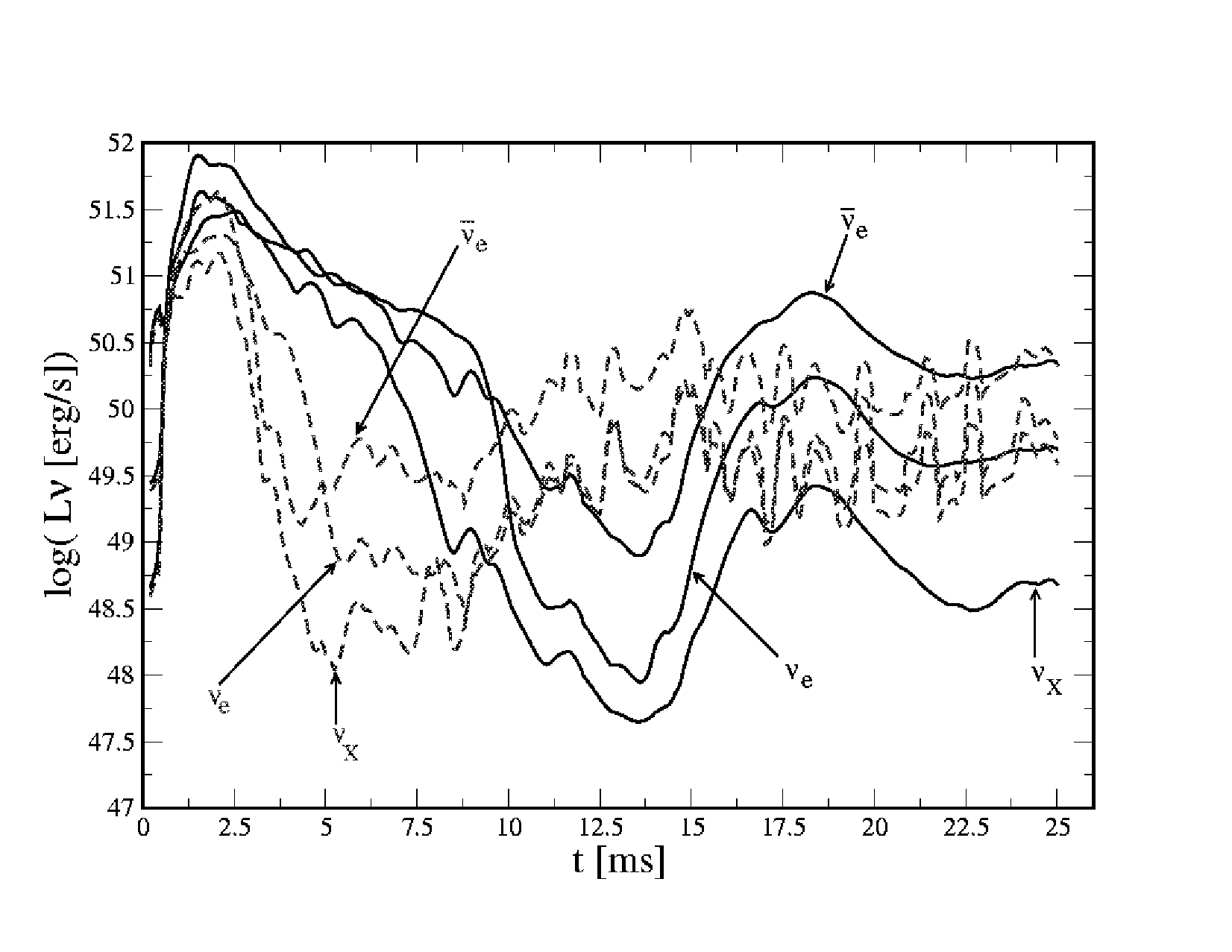}}
  \caption{Neutrino luminosities of the neutron star black hole cases. Left
  panel: black holes with 3 (dashed) and 4 \msun (solid curves). 
  Right panel: black holes with 7 (solid) and 10 \msun (dashed curves).}   
  \label{fig:nu_lum_nsbh}
\end{figure*}
Similar to the case of a new-born proto-neutron star,
the huge neutrino luminosity will also drive an energy baryonic winds off 
the remnant. The interaction with the baryonic
material collimates the neutrino-annihilation-driven bipolar outflow. Using
simple estimates Rosswog and Ramirez-Ruiz (2003) found a broad 
distributions of opening angles and isotropised luminosities, centered 
around 7$^\circ$ and a few times $10^{50}$ erg, respectively.\\
More recently, Aloy et al. (2005) have performed detailed, axisymmetric 
relativistic hydrodynamic simulations in which they parametrised the energy 
deposition above remnant accretion disks. Some of
their models did not produce ultra-relativistic outflows and instead lead to
low-luminosity UV-flashes, others were able to produce ultra-relativistic, 
structured outflows that are promising GRBs sources. While the
ultimate outcome seems to depend quite sensitively on the details of the
considered system (disk structure, energy deposition rates, injection
duration), their most promising systems produced ultra-relativistic
outflows with half-opening angles of 5-10$^\circ$ and apparent isotropic
energies of up to $10^{51}$ erg.\\
%confrontation with reality
It has to be stressed that to date no self-consistent calculations of a full
merger with neutrino emission, annihilation, launch of winds and relativistic
outflow could be performed. Only individual aspects have been simulated that
are a posteriori patched together into a global picture. Nevertheless, the 
theoretical numbers found in the last two investigations are consistent with
recent observations of short GRBs that find values of $E_\gamma$ of a 
few times $10^{48}$ erg and opening half-angles from about 9 to about 
14$^\circ$ (Fox et al. 2005, Berger et al. 2005). Note however, that there 
are recent observations (Berger et al. 2006b) that suggest that a fraction of 
short GRBs may have substantially larger energies than the first few short 
bursts that were detected.\\
Thus, neutrino annihilation from compact merger remnants, at least in the DNS
case, seems to be in reasonable agreement with recent observations of short 
GRBs.\\

\subsection{Magnetic fields}
\begin{figure}[!t]
  \includegraphics[angle=-90,width=0.95\columnwidth]{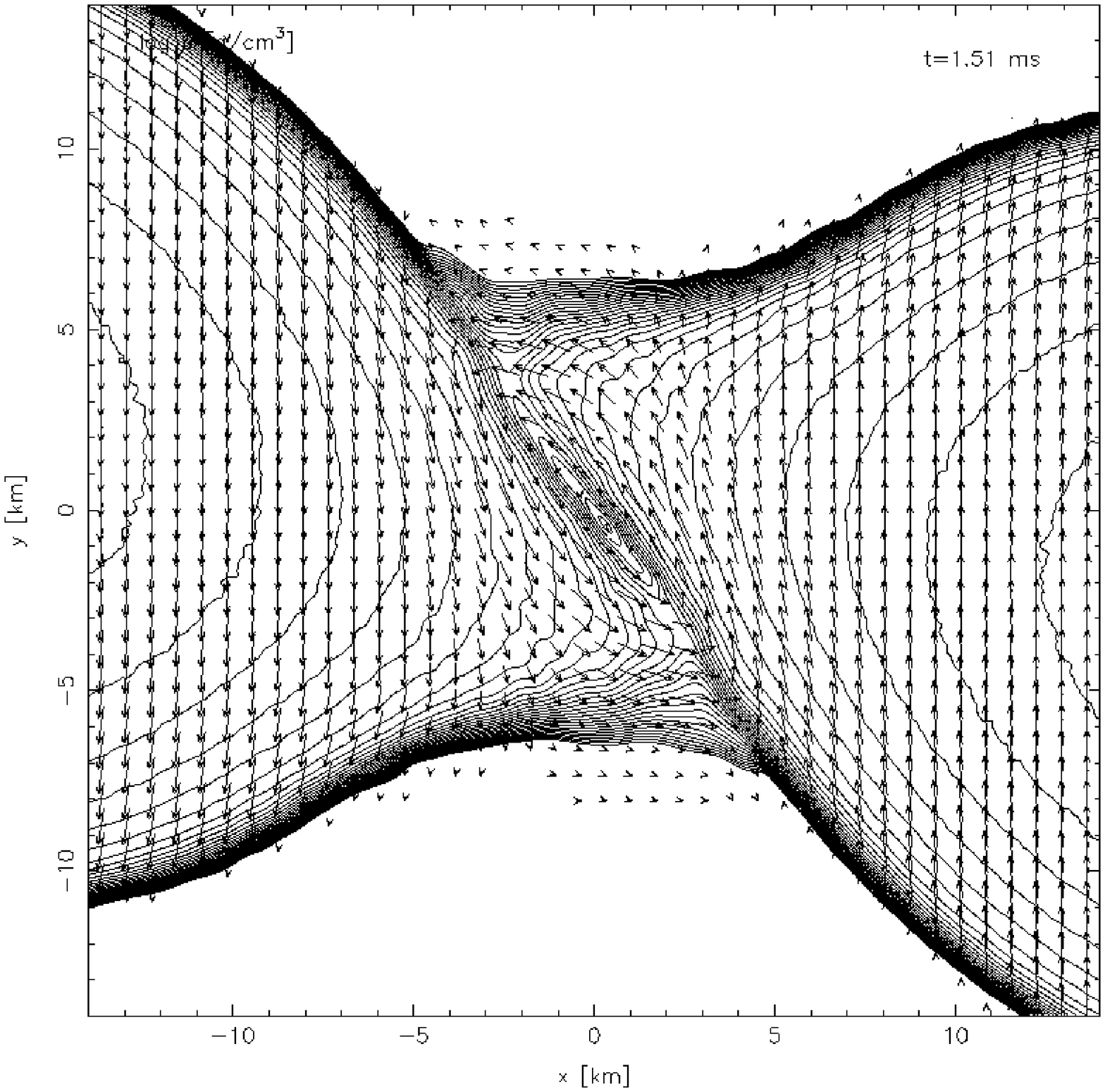}
  \includegraphics[angle=-90,width=0.95\columnwidth]{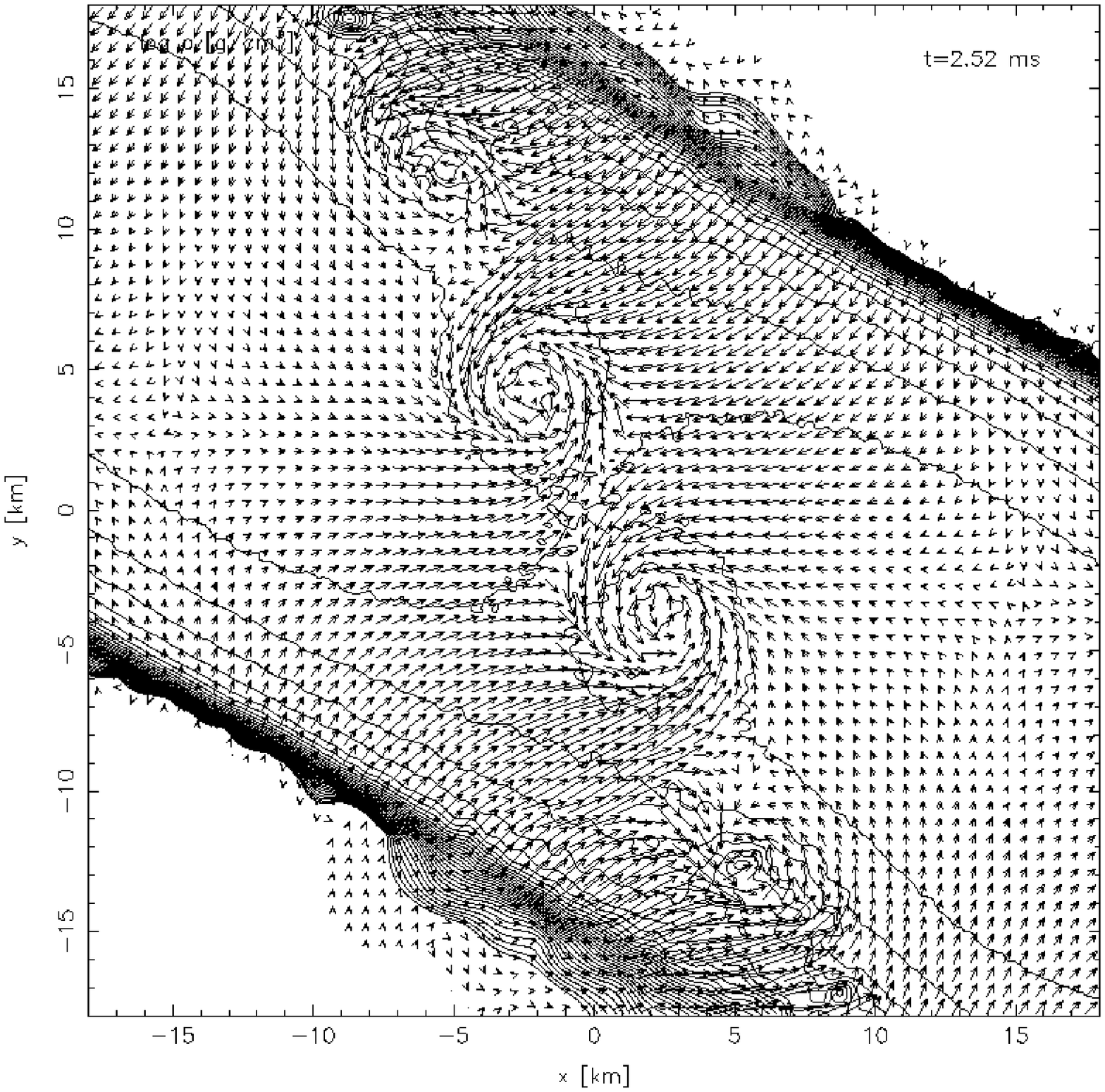}
  \caption{Velocity fields during the merger of two 1.4 \msun neutron
  stars (run DA). The shear interface, see panel one, becomes 
  Kelvin-Helmholtz unstable and forms a string of vortex rolls 
  (panel two, velocities in corotating frame) in which the magnetic field is
  curled up.}   
  \label{fig:shear_KH}
\end{figure}

Nevertheless, it is hard to see how magnetic fields should not play an
important role. Most neutron stars are endowed with strong magnetic fields:
young pulsars typically have $\sim 10^{12}$ G and  magnetars (Duncan and 
Thompson 1992)  are thought to have magnetic fields between
$10^{14}$ and $10^{15}$ G. While still more speculative than
the neutrino annihilation mechanism in the sense that the existing
calculations are less detailed, there are good  arguments to expect a
dramatic growth of existing seed fields.\\
The merger debris is differentially rotating, both the central object before
collapse, see Fig.~9 in Rosswog and Davies (2002), and the surrounding disk.
Thus the initial seed fields can grow either just by a linear winding up the
field lines or, more likely, via the magnetorotational instability (MRI)
(e.g. Balbus and Hawley 1998). Recent magnetohydrodynamics simulations (Price
and Rosswog 2006) suggest that magnetic seed fields are rapidly amplified in
the Kelvin-Helmholtz instability at the shear interface between the two
neutron stars, see Fig.~\ref{fig:shear_KH}. This amplification occurs on a
timescale of only $\sim$ 1 ms, long before the collapse 
can set in. As the shortest modes grow fastest in a Kelvin-Helmholtz
instability, the reached maximum field strengths are limited by the finite
numerical resolution. At the highest affordable resolution, about
$2\cdot10^{15}$ G are reached, but very plausibly {\em much} higher 
field strengths
will be realized in nature. The kinetic energy of the central object is large,
about $8\cdot10^{52}$ erg, and if say 10 \% of this energy can be transformed
into magnetic field energy, the field strength averaged over the central
object will be in excess of $10^{17}$ G. Locally, the field could be
substantially higher.\\
\begin{figure}[!t]
  \includegraphics[angle=-90,width=1.1\columnwidth]{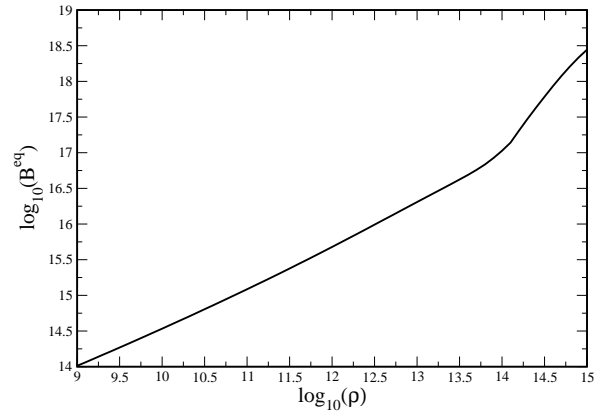}
  \caption{Field strength (in G), where the magnetic pressure, 
           equals the matter pressure (calculated from the Shen et
           al. EOS (1998) with $T=0.5$ MeV and $Y_e=0.1$).}   
  \label{fig:Beq}
\end{figure}
To become dynamically important, the fields have to become extremely strong.
The field strengths at which the magnetic pressure balances the nuclear matter 
pressure are shown in Fig.~\ref{fig:Beq}. In the central object of a merger 
these field strengths are in excess of $10^{17}$ G, in the accretion disk 
they are still between $10^{14}$ and $10^{16}$ G.
%possibilities: i) superpulsar 
If the central object remains stable for long enough, it could provide a
Poynting-dominated, scaled-up relativistic pulsar wind. In this case 
energy is released at a rate of 
\be
\left(\frac{dE}{dt}\right)_{\rm md} \sim 10^{51} B_{16}^2 P_{\rm ms}^{-4} R_6^6
\; {\rm erg/s}
\ee
and the central object will spin down on a timescale given by
\be
\tau_{\rm sd} \sim  2 {\rm s} \; B_{16}^{-2} P_{\rm ms}^2 \left(\frac{20 \; {\rm
      km}}{R_{\rm co}}\right)^4  \left(\frac {M_{\rm co}}{2.5 \; {\rm M}_\odot}\right),
\label{eq:dipole_spindown}
\ee
where $B_{16}$ is $B/10^{16}$G and $P_{\rm ms}= P/$1 ms.

%ii) buoyancy 
Fluid instabilities, such as the Kelvin-Helmholtz 
instability illustrated for run DA in Fig.~\ref{fig:shear_KH}, will locally
curl up the magnetic field lines. This had been suggested earlier
(Rosswog et al. 2003) and has recently been confirmed by first MHD-merger
simulations (Price and Rosswog 2006). Once such high-field pockets reach field
strengths close the equipartition, see Fig.~\ref{fig:Beq}, they will become
buoyant, float up and produce explosive reconnection events (e.g. Kluzniak and
Ruderman 1998, Rosswog et al. 2003). This may also occur in the disk
(e.g. Narayan et al. 1992), though at a lower field strength, see
Fig.~\ref{fig:Beq}. 
%iii) if black hole exists:
%magnetic coupling to the disk -> extraction of black hole spin energy
Once the central object has collapsed, the ``standard
GRB-engine'', a black hole plus accretion torus, remains, for which various
energy extraction mechanisms have been suggested  (e.g. Blandford and Znajek
1977, Blandford and Payne 1982 or McKinney 2005). For a discussion of these 
mechanisms we refer to the literature.\\
As discussed in the previous sections in our neutron star black hole
calculations we find it difficult to form massive disks. The neutrino
luminosities are even in the more optimistic cases two orders of magnitude
lower than in the DNS case, see Figs.~\ref{fig:nu_lum_DNS} and
\ref{fig:nu_lum_nsbh}. This is mainly due to the much smaller debris mass, see
Fig.~\ref{fig:bound_debris}. As the neutrino annihilation roughly scales with
the square of the neutrino luminosity, see Eq.~(\ref{eq:ann}), neutrino
annihilation deposits substantially less energy than in the DNS case.  
NSBH binaries would therefore not produce energetic GRBs, but rather a 
low-luminosity tail to the short GRB distribution. Possible observable 
signals are discussed in Rosswog (2005).\\
We want to stress again that these results are sensitive to the nuclear 
EOS. With the softer Lattimer-Swesty-EOS Janka et al. (1999) find more 
promising conditions.

\section{Formation of a magnetar?}
\label{sec:magnetar}
There are essentially two proposed formation mechanisms for pulsars: i) the
collapse of very rapidly rotating stars with ordinary magnetic fields or ii)
the collapse of extraordinarily magnetized stars with ordinary rotational
speeds. In the first case the new-born proto-neutron star has to be
rotating rapidly enough so that rotational periods are smaller than the
overturn times, $\sim 3$ ms, of convective eddies (Duncan and Thompson 1992,
Thompson and Duncan 1993, Thompson and Duncan 1995), so that an efficient
$\alpha$-$\omega$-dynamo can become active. In this way, fields as strong as
$\sim 10^{15}$ G can be plausibly generated. Such a new-born neutron star will
be spun down quickly by magnetic dipole radiation (and possibly mass loss) and
this will fuel the 
supernova explosion with an additional $2 \pi^2 I/P^2\sim 2 \cdot 10^{52}
I_{45}/P_{\rm ms}^2$ ergs stemming from rotational energy. Here we have used
$I_{45}$, the moment of inertia in units of $10^{45}{\rm g cm}^2$ and $P_{\rm
 ms}$, the rotation period as measured in milliseconds. Therefore,
particularly bright supernova explosions going along with magnetar formation
are a natural prediction of this formation path. However, the three supernova
remnants that seem to be associated with magnetars are remarkably 
unremarkable: rather than being particularly energetic they only have 
kinetic energies of $\sim 10^{51}$ ergs (Sasaki et al. 2004, Vink 2006). 
An alternative formation scenario could be that the magnetar field 
strengths are obtained via magnetic flux conservation alone during the 
collapse of highly magnetized progenitor stars (Ferrario and 
Wickramasinghe 2005).\\
Here, we want to discuss a third possible formation channel, namely 
a neutron star coalescence that produces produces a stable, magnetar-like 
remnant. 

\begin{figure}
{\includegraphics[angle=0,width=0.75\columnwidth,angle=-90]{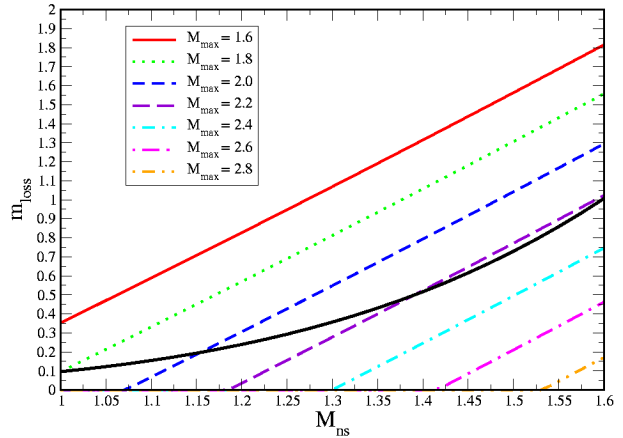}}
\caption{Shown is the baryonic mass that has to be lost from a double neutron
  star system with twice $M_{ns}$ if the maximum gravitational neutron star
  mass is $M_{\rm max}$. All masses are in solar units.
  The black solid curve is the estimate of the mass that is lost by the
  neutrino-magnetocentrifugal wind plus dynamic ejection, see text.
 \label{fig:baryonic_mass}
} 
\end{figure}

% rates
The first point to clarify is whether the DNS merger rates are large enough to
possibly contribute a non-negligible fraction of magnetars.
The estimated rates at which double neutron systems (DNS) merge have been
constantly increasing over the last years. The discovery of the double pulsar
PSR J0737-3039A (Burgay et al. 2003) alone has increased the estimated merger
rate based on observational data by an order of magnitude. Current estimates
(Kalogera et al. 2004) are in the range from 4 to 224 $\cdot10^{-6}$ per year
and Galaxy. Nakar et al. (2005) based their estimates on the
assumption that short-hard gamma-ray bursts result from DNS mergers. They find
a ``best guess'' value that is higher than the above upper limit by more than
an order of magnitude. This last rate may be somewhat optimistic, but it is
fair to state that the DNS merger rate is comparable to the magnetar 
birthrate in our Galaxy, about 1-10 $\cdot10^{-4}$ per year 
(Duncan and Thompson 1992). DNS mergers could therefore plausibly contribute to
the magnetar production.\\
%maximum neutron star mass
The main question that needs to be addressed is whether the final remnant of 
a DNS merger can have a mass that can be permanently sustained by the
equation of state. The maximum neutron star mass is unfortunately still only
poorly constrained. An upper limit based on very general physical 
principles such as causality is 3.2 \Msun (Rhoades and Ruffini 
1974)\footnote{See Psaltis (2004) for a critical analysis
of the underlying assumptions.}. In a review of realistic models of 
nuclear forces
Akmal et al. (1998) find that realistic models of nuclear forces limit the
maximum mass of neutron stars to be below 2.5 \Msun. The masses
with the smallest error bars come from the observation of DNS and they
are consistent with a remarkably narrow underlying Gaussian mass distribution
with M= 1.35 $\pm$ 0.04 \msun (Thorsett and Chakrabarty 1999). Observations of
neutron stars in binary systems with a white dwarf, however, yield 
consistently higher neutron star masses, but the error bars are to date 
still much larger than in DNS. The most extreme case is PSR J0751+1807 
with an estimated pulsar mass of $2.1\pm 0.2$ \msun (Nice et al. 2005). 
Recent observations of the neutron star EXO 0748-676 also seem to support 
large possible neutron star masses and have been interpreted as hints 
that neutron stars may have a conventional neutron-proton composition 
(\"Ozel 2006). Here, we want to explore the possibility that the final 
outcome of a DNS coalescence is, after some mass loss, a very massive and
highly magnetised, but stable neutron star.\\ 
%Mass loss
% How below maximum mass ?
% i) gravitational vs. baryonic mass -> Yahil Lattimer
% ii) Mass loss: a) dynamic
%                b) neutrino wind
%                c) magnetic
Let us first estimate the gravitational mass of a merger remnant.
Lattimer and Yahil (1989) found empirically that 
\be
B= \alpha \left( \frac{M}{M_{\odot}}\right)^2,  
\ee
with $\alpha=0.084$ \Msun, yields a good fit of the binding energy, $B$, as a
function of the gravitational mass $M$. Thus, a standard binary system of
twice 1.35 \msun corresponds to a baryonic mass of about 3.0 \Msun, or, a
merger remnant made of all the baryons of the initial binary system would have
2.48 \msun of gravitational mass.\\
We plot in Figure \ref{fig:baryonic_mass} the baryonic mass that needs 
to become unbound in order for the final remnant to be 
above a specified maximum neutron star mass M$_{\rm max}$:
\be
m_{\rm loss}= 2(\alpha M_{\rm ns}^2 +  M_{\rm ns}) -  (\alpha M_{\rm max}^2 +  M_{\rm max}),
\ee
where we have restricted ourselves to the case of equal mass neutron
stars. For example, if we 
assume the maximum neutron star mass to be 2.2 \Msun, so just beyond the
estimated mass of PSR J0751+1807, a system with two 1.2 (1.4) \msun neutron
stars would need to loose a baryonic mass of 0.035 (0.52) \Msun.\\
Mass can be lost by dynamical ejection or by the combined action of neutrinos,
rotation and strong magnetic field as a neutrino-magnetocentrifugally driven 
outflow. The latter mechanism has been discussed recently (Thompson et al. 
2004) in the context of the formation of magnetars in supernovae.\\
The amount of mass that is {\em dynamically} ejected is sensitive to the 
nuclear equation of state (EOS) with stiffer EOSs ejecting more than 
softer ones (Rosswog et al. 2000). With the EOS of Shen et al. (1998), 
which is at the stiffer end of the EOS spectrum and therefore consistent with 
the assumption of a relatively large maximum neutron star mass, we find that 
typically $m_{ej}\approx 0.03$ \msun are dynamically ejected.\\
The combined action of neutrino-heating, strong magnetic field and rapid
rotation is hard to quantify as details of the 
(unknown) magnetic field geometry enter and the mass loss rates depend
sensitively on the neutrino luminosity and the rotational period (Duncan et
al. 1986, Qian and Woosley 1996; Thompson et al. 2001 and Thompson et
al. 2004). 
% Thompson-results
Strong magnetic fields enforce near-corotation of the wind with the stellar
surface out to the Alfven radius, $R_A$, where the kinetic enery, $\rho v^2/2$
approximately equals the magnetic energy density $B^2/8\pi$. Thus, wind mass
elements are, like spokes on a bike wheel, forced to corotate with the
remnant, leading much faster loss of angular momentum and spin-down. Thompson
et al. (2004) find that a strong magnetic field together 
with the rapid rotation will {\em drastically} increase the mass loss rate, in
the fast-rotation limit they find a functional dependence of
\be
\dot{M} \propto \exp(\omega^2),
\ee
where $\omega$ is the angular frequency. This indicates that a substantial
mass loss is possible.
For example, for a central remnant period of 0.8 ms (see Figure 10 in 
Rosswog and Davies 2002)
and a neutrino luminosity of $L_{\bar{\nu}_e}= 8\cdot 10^{51}$ erg/s they find
a mass loss rate $\dot{M} \approx 0.1$ \msun/s. If we assume that the results
scale like $\dot{M} \propto L_{\nu}^{2.5} M^{-2}$ (Qian and Woosley 1996) and
use the empirical relation Eq.~(\ref{eq:lnu_tot}), we can estimate $\dot{M}$ 
as a function of the total mass of the DNS, $M$. This is shown as the solid 
black line in Fig. \ref{fig:baryonic_mass} for a duration of $\tau= 0.5 s$, 
the total unbound mass being $m \sim \dot{M} \cdot \tau + m_{ej}$. Taking 
these numbers at face value would mean that, if the maximum neutron star 
mass is $M_{\rm max}=2.2$ \msun (violet line in Fig.\ref{fig:baryonic_mass}), 
then even a standard DNS with twice 1.35 \msun would loose enough
mass to leave a stable remnant. For $M_{\rm max}= 2.0$ \msun (blue), 
neutron stars with or below 1.15 \msun would still produce a stable 
neutron star.\\
To summarize, although hard numbers are difficult to obtain, 
the lower end of the DNS mass distribution could plausibly produce stable, 
highly magnetized neutron stars rather than black holes.\\ 

\section{Late-time central engine activity}
\label{sec:late_signals}
It was one of the surprises of the recent afterglow detections of short GRBs
that GRB 050709 and GRB 050724 exhibited long-lasting X-ray flares that
occurred at least 100 s after triggering the bursts. While the observations
have been claimed to be suggestive for compact mergers, it is not immediately
clear whether the merger model can accommodate this late time X-ray flaring
activity. The standard central engine, black hole plus disk, has --at least in
its most simple form-- difficulties to accommodate these long time scales.
The dynamical timescale for an accretion disk is
\be
\tau_{\rm d} \sim \frac{1}{\alpha \Omega_K} \sim 0.05 {\rm s} 
\left(\frac{R}{200\; {\rm km}}\right)^{\frac{3}{2}}\left(\frac{0.1}{\alpha}\right)\left(\frac{2.5
    M_{\odot}}{M_{\rm bh}}\right)
\ee
where $\alpha$ is the Shakura-Sunyaev viscosity parameter (Shakura and Sunyaev
1973) and $\Omega_K$ the angular frequency of a ring of matter at distance
R. Obviously, the estimated accretion timescale is much shorter than the
timescales at which the X-ray flaring activity is observed. \\
The X-ray activity might be caused by either a long-lived central engine, 
or by a wide distribution of Lorentz factors, or else by the deceleration of a
Poynting flux dominated flow. Various possibilities are discussed in detail
in Zhang et al. (2005). Here we want to address two possibilities:
i) that the central object from a DNS merger survives for long enough to
produce the flaring and ii) we also address the fallback accretion luminosity
in the aftermath of a compact binary merger.\\ 
%possibilities: i) central engine keeps going, ii) fallback or iii) something
%else not related to engine

\subsection{Survival of the central object in a double neutron star merger}
\label{sec:survival}
The continued central engine activity could be due to 
the survival of the central object, either in a meta-stable state or --as
discussed above-- the remnant may in some cases loose enough mass to produce
a stable, highly magnetised neutron star.\\
A meta-stable state is likely to occur, stabilisation could come for example
from differential rotation (e.g. Morrison et al. 2004) or trapped neutrinos 
together with non-leptonic negative charges (Prakash et al. 1995), but the
time scale until collapse is uncertain. This is because
all the imprecisely known aspects of the problem such as the EOS at 
supra-nuclear densities, the remnant radius and magnetic field, ellipticity 
of the central object and so on, enter with large powers in the estimate of 
the time scale. Therefore in the meta-stable case a large range 
of time scales until collapse can be expected for different initial binary
systems. With reasonable parameters even time scales as long as weeks are possible
(see Rosswog and Ramirez-Ruiz 2002).\\
If the GRB is indeed caused by the combined action of neutrino annihilation
and buoyant pockets of ultra-strong magnetic fields, this will plausibly also
cause late-time flaring activity, provided that the remnant is stable for long
enough. The magnetic field amplification acts on a time scale of about a
milli-second (Price and Rosswog 2006) and thus large magnetic fields can be
built up before the neutrino luminosity has reached its peak at $\approx 20$
ms after the merger, see Fig.~\ref{fig:nu_lum_DNS}. In the early stages 
a buoyant high-field bubble can  expand into a practically baryon-free
environment and thus reach large Lorentz-factors. At neutrino
luminosities in excess of $10^{52}$ erg/s the remnant also drives a very
energetic baryonic wind. If the right conditions are met, the baryonic
material can help to collimate the outflow (Aloy et al. 2005), but
it also poses a potential threat to the emergence of an ultra-relativistic
outflow. Initially, as the rapidly spinning remnant produces high-field
pockets in rapid succession, magnetic pressure may help to keep the funnel
above the central object baryon-clean. But as the remnant is constantly braked
by magnetic fields and gravitational wave emission, it takes
longer and longer to reach buoyancy field strength, while more and more
baryons are ablated from the remnant by neutrinos. The first explosive reconnection events
escape into a very clean environment, those produced later will be released 
into an envelope of neutrino-magneto-centrifugally expelled baryons. This 
different baryon loading may be responsible for the difference between 
bursts and X-ray flares.

\subsection{Fallback accretion}
Some of the matter expelled by gravitational torques is still gravitationally
bound to the central object, and will  fall back towards the remnant (see for
example Figure \ref{fig:DNS_1.1_1.6}, panel three). Direct numerical
calculations over the interesting timescales are not feasible, therefore we
follow a simplified analytical approach. We assume that the motion of this
material can be approximated as Keplerian motion in a central potential
produced by the remnant. From the total energy, $E_i$, and the angular
momentum, $J_i$, of each SPH-particle the eccentricity of the particle orbit,
$e_i$, can be calculated 
\be
e_i^2= 1 + \frac{2 E_i J_i^2}{G^2 m_i^3 M^2}.
\ee
Here, $m_i$ is the particle and $M$ the enclosed mass. For the particles with
$e_i<1$, the semi-major axes
are calculated, $a_i= -G  m_i M/(2 E_i)$, from which the distances of the 
closest and farthest approach follow: $R_{\rm min}= a_i (1-\epsilon_i)$ and
$R_{\rm max}= a_i (1+\epsilon_i)$. A particle with a velocity $\vec{v}_i$
that is currently located at radius $r_i$ will reach the radius $R_{\rm dis}$ 
after a time of 

\be
\tau_{i}=
\left\{\begin{array}{cl}
I_{r_i,r_{{\rm max},i}} + I_{r_{{\rm max},i},R_{\rm dis}}  
\quad {\rm for} \quad \vec{v}_i\cdot\vec{r}_i > 0\\
I_{r_i,R_{\rm dis}} \hspace*{2.5cm} {\rm for} \quad 
\vec{v}_i\cdot\vec{r}_i < 0
\end{array}\right. \label{fallback_time},
\ee
where $I_{r_1,r_2}$ is given by

\begin{eqnarray}
I_{r_1,r_2}=&&
\left[ \frac{\sqrt{A r^{2}+ B r + C}}{A} \right. \nonumber \\
&+& \left. \frac{B}{2A
    \sqrt{-A}} \arcsin \left( \frac{2 A r + B}{\sqrt{-D}} \right)
\right]_{r_1}^{r_2}
\label{I_ri_R}
\end{eqnarray}
with $D = 4 A C - B^{2}$ and $A= \frac{2 E_{i}}{m_{i}}, \; B= 2 G M$ and 
$C=-\frac{J_{i}^{2}}{m_{i}^{2}}$ (Rosswog 2006).

\begin{figure}
\hspace*{-0.4cm}\psfig{file=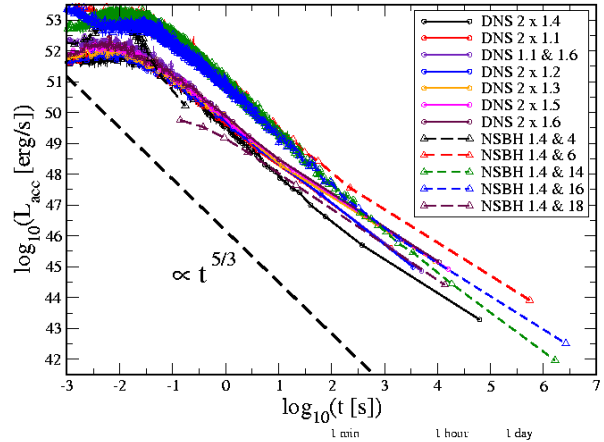,width=0.95\columnwidth,angle=-90} 
\caption{Fallback accretion luminosity for various compact binaries, taken
  from Rosswog (2006).
Circles refer to double neutron star systems (DNS), triangles to neutron star
black hole mergers (NSBH). Note that the rightmost point is determined by the 
mass resolution of the simulation. For reference a straight line with slope
5/3 is shown.}
\label{fig3}
\end{figure}

For the radius, where the fallback energy is dissipated, $R_{\rm dis}$, 
we take in the double neutron star case the disk radius at the end of the 
numerical simulation. This is conservative in the sense that the disk will
shrink, see below, and therefore the assumption of a radius fixed at 
$R_{\rm dis}$ yields shorter timescales and lower accretion luminosities. 
For the black hole cases, I choose $R_{\rm dis}= 10 GM/c^2$, so just outside 
the innermost stable circular orbit of a non-rotating (Schwarzschild-) black
hole at $R_{\rm ISCO}= 6GM/c^2$. The details of the fallback times and energies
change slightly with $R_{\rm dis}$, but none of the conclusions
depends on the exact numerical value of $R_{\rm dis}$.\\
Fig.~\ref{fig3} shows the accretion luminosities, 
$L_{\rm acc}= dE_{\rm fb}/dt$, derived for various DNS and NSBH systems, for
details see Rosswog (2006).
Here, $E_{\rm fb}$ denotes the difference between the potential
plus kinetic energy at the start radius, $r_i$, and the potential energy at
the dissipation radius, $R_{\rm dis}$. The curves have been obtained 
by binning the energies contained in the fallback material, $E_{\rm fb}$,  
according to the corresponding fallback times, $\tau_{i}$, see 
Eq.~(\ref{fallback_time}). A fraction $\epsilon$ of this energy is channeled 
into X-rays, $L_X = \epsilon L_{\rm acc}$.\\
The double neutron star cases are rather homogeneous with respect
to their fallback accretion, in all cases the fallback material is
approximately 0.03 \Msun\footnote{General relativistic effects might reduce
  this number somewhat.}. After an initial, short-lived plateau the luminosity
falls off with time close to the expected 5/3-power law
\citep{rees88,phinney89}.  
The last point in these curves is determined by the numerical mass 
resolution in the hydrodynamics simulations and should therefore be 
interpreted with some caution. All other points should be a fair 
representation of the overall fallback activity. Typically, the X-ray 
luminosity about one hour after the coalescence is 
$L_X  \sim \left(\frac{\epsilon}{0.1}\right) \cdot 10^{44}$ erg/s.
For the investigated mass range the spread in the luminosities one hour after
the coalescence is about one order of magnitude.\\
The neutron star black hole cases show a larger diversity. The mass in the
fallback material of different mass ratios varies by about a factor of 500,
see Rosswog (2006), an hour after the merger the accretion 
luminosities of the different NSBH systems differ by about two orders of 
magnitude. The involved time scales change strongly with the binary mass
ratio. For example, the 1.4 and 4 \msun NSBH case does not produce much
eccentric fallback material. Accretion, at least to the resolvable level, is
over in $\sim 0.2$ s. This accretion period may produce a short GRB, but
probably not much X-ray activity. The 1.4 and 18 \msun NSBH system is at the
other extreme: its peak luminosity is lower by three orders orders of
magnitude but extends (at a resolvable level) up to about one hour. The mass
ratios in-between could possibly produce a (weak) GRB and extended X-ray
activity up to about one day after the burst. 

\section{Summary}
%GWs
We have discussed various aspects of the last stages in the life of a 
compact binary. In particular we have addressed

\bi
\item Gravitational waves as a probe of the merger dynamics

We have shown how the merger dynamics is imprinted on the gravitational wave
signal. Examples include the peak gravitational wave frequency of a 
double neutron star merger and  the subsequent ringdown emission which both
are sensitive to the equation of state. For a stiff EOS, low-mass 
black hole neutron star systems undergo a long-lived, mostly episodic 
mass transfer phase. We show an example where the mass transfer 
continues for as long as 47 orbital periods. This last binary phase is 
responsible for a gravitational wave signal with slowly decaying amplitude 
and increasing frequency. The signal only shuts off when the
neutron star is finally disrupted upon reaching its minimum mass.\\
The coincident of detection of a short GRB together with a ``chirping''
gravitational wave signal would be the ultimate proof of a compact binary
central engine.

\item Accretion disks

Compact binary mergers produce in some (but not all) cases
accretion disks from the neutron star debris. These disks differ from the
standard thin disk accretion model in various ways. Due to insufficient
cooling by neutrinos they are often thick and puffed up, they are usually
neither in a dynamical nor in $\beta$-equilibrium. For double neutron star
mergers they are generally massive, of order 0.25 \Msun, with temperatures of
several MeV and an electron fraction of close to 0.1 (at least initially). 
For neutron star black hole systems we find it substantially more difficult to
form massive disks. For low-mass black hole systems, the initial episodic mass
transfer with a surviving neutron star prevents the build up of a massive
disk. Only after a complete disruption of the neutron star a disk of $\approx
0.05$ \msun forms. A comparison of the debris masses in double neutron star
and neutron star black hole mergers is shown in Fig.~\ref{fig:bound_debris}.
We find the neutron star black hole systems to produce disks masses that 
are lower by at least a factor six. For higher mass black holes the neutron 
star is disrupted early on, but very close to the innermost stable circular
 orbit. The resulting accretion disks are geometrically thin and essentially 
cold. For black hole masses larger that about 16 \msun no disk forms at 
all, most of the neutron star is fed directly into the hole, the rest is 
dynamically ejected.\\
These results are sensitive to the equation of state, for a softer equation of
state the neutron star is disrupted more easily. 

\item Gamma-ray bursts
  
%nu-annihilation
Neutrino annihilation from the remnant of a double neutron star
merger can plausibly provide the driving stresses to launch highly
relativistic, bipolar outflows, but a large diversity in the observable burst
properties and moderate burst energies of $\sim 10^{48}$ are expected. 
It is conceivable that a good fraction of systems
fails to provide the right conditions and instead produces X-ray or
UV-flashes.\\
MHD simulations of double neutron star mergers show that the magnetic field
grows within the first millisecond to (probably much) beyond magnetar
strength, possibly allowing for a burst production $\grave{\rm a}$ la 
Kluzniak and Ruderman (1998). The burst is most likely a result of the 
combined action of both neutrino annihilation and ultra-strong magnetic 
field.\\
 Of the neutron star black hole binaries only systems with low black
hole masses could possibly form disks that are hot and dense enough to launch 
GRBs (unless the true nuclear equation of state is much softer than the one we
use here). Among them, rapidly spinning holes are preferred as both the last
stable orbit and the event horizon move closer to the hole. 
The high-mass end of neutron star black hole binaries probably produces a
low-luminosity tail of the short GRB-distribution.
%magnetic fields
%The magnetic fields of the neutron stars are expected to be amplified during
%the merger, and probably drastically so. Recent calculations for double
%neutron stars starting with a
%normal pulsar field of $10^{12}$G indicate that field strength beyond magnetar
%strengths could be reached within the first millisecond after contact. If
%stable for long enough, instabilities will locally amplify the field until it
%becomes buoyant and dissipates in flares above the remnant. This is expected
%to produce an erratic sequence of blasts. The surrounding will initially be
%very baryon-poor, but subsequently be loaded by baryons ablated by the large
%neutrino emission. Later blasts will therefore have to cope with much
%larger baryon loadings.\\
%
\item Magnetar formation

If the maximum neutron star mass should indeed be as high as indicated by
recent observations the amount of mass that would need to be lost 
after a merger to evade the final collapse to a black hole
is not implausibly large. Therefore we want to point out the possibility
that binary systems from the lower end of the mass distribution could 
produce a highly magnetised, but stable neutron star rather than a black 
hole. Magnetars outside normal supernova remnants
would be obvious candidates for this formation channel. If true, this channel 
should also produce large amounts of neutron-rich nuclei synthesised in 
both the dynamically ejected neutron star material and in the 
neutrino-magnetocentrifugal wind that made the survival of the remnant
possible in the first place. Being synthesised in a neutrino bath dominated
by electron anti-neutrinos the wind material will to be more
proton-rich than the dynamically ejected material. The exact $Y_e$ is
determined by both the expansion time scale and the ratio of neutrino and
anti-neutrino luminosities, but is roughly (Qian and Woosley 1996)
\be
Y_e \sim 0.28 \left[1 + \frac{L_{\rm \bar{\nu}_e}/
5 \cdot 10^{52} {\rm erg}}{L_{\rm \nu_e}/
2 \cdot 10^{52} {\rm erg}}\right]^{-1},
\ee
where $L_{\rm \nu_e}$ and $L_{\rm \bar{\nu}_e}$ are the electron neutrino
and anti-neutrino luminosities, respectively. The more 
neutron-rich dynamical ejecta with $Y_e\approx 0.1$ would be expected to 
be located at larger distances from the magnetar than the wind material.  

\item Late time X-ray activity

In the speculative case that the central object of a neutron star 
merger survives, late-time flaring would be caused by the same 
mechanism that produced the burst. If the burst is (at least in part) produced
by explosive reconnection events of buoyant high-field bubbles as suggested
by Kluzniak and Ruderman (1998), parts of the remnant that become buoyant at
later times have to expand into an environment that is already heavily
baryon-polluted by neutrino-driven winds. Due to the higher baryon loading
the emission would be mainly in the X-ray band of the spectrum.

Much less speculative is the accretion of fallback material which occurs
naturally in the aftermath of a compact binary merger. A few percent of 
a solar mass are ejected into eccentric, but still bound orbits. The 
fallback of this material produces extended X-ray activity up to many hours 
after the main burst.
\ei

\acknowledgements
\noindent Some of the figures have been produced using the 
software SPLASH kindly provided by Daniel Price.\\
The calculations shown in this paper were in part performed on the JUMP system
of the H\"ochstleistungsrechenzentrum J\"ulich.\\
It is a pleasure to thank Joe Monaghan and Andrew Melatos for their
hospitality during the last stages of writing this paper.

%\bibliography{astro_SKR}
%\bibliographystyle{mn2e.bst}

\end{document}